\documentclass[nofootinbib,amsmath,amssymb,prd,notitlepage,superscriptaddress]{revtex4-2}

\pdfoutput=1

\usepackage{color}
\usepackage{graphicx}
\usepackage{dcolumn}
\usepackage{bm}
\usepackage{bbm}
\usepackage{amsmath}
\usepackage{amssymb}
\usepackage{mathtools}
\usepackage{float}
\usepackage{physics}
\usepackage[a4paper, total={6.5in, 9.5in},margin=1in]{geometry}

\def\hhref#1{\href{http://arxiv.org/abs/#1}{arXiv:#1}} 
\usepackage{hyperref}

\DeclareMathAlphabet{\mathdutchcal}{U}{dutchcal}{m}{n}
\SetMathAlphabet{\mathdutchcal}{bold}{U}{dutchcal}{b}{n}
\DeclareMathAlphabet{\mathdutchbcal}{U}{dutchcal}{b}{n}

\begin{document}

\title{On the Higher Loop Euler-Heisenberg Trans-Series Structure}

\author{Gerald V. Dunne and Zachary Harris\\
Department of Physics, University of Connecticut, Storrs, CT 06269-3046, USA}

\begin{abstract}
We show that the one-loop Euler-Heisenberg QED effective Lagrangian in a constant background field acquires a very different non-perturbative trans-series structure at two-loop and higher-loop order in the fine structure constant. Beyond one-loop, virtual particles interact, causing fluctuations about the instantons, whereby the simple poles of the one-loop Borel transform become branch points. We illustrate this in detail at two-loop order using Ritus's seminal result for the renormalized two-loop effective Lagrangian as an exact double-integral representation, and propose a possible new approach to computations at  higher loop order. Our methods yield remarkably accurate extrapolations from weak-field to strong-field, and from magnetic to electric background field, at both one-loop and two-loop order, based on surprisingly little perturbative input.

\end{abstract}

%\date{\today}

\maketitle

\section{Introduction}

\begin{figure}[t]
\centering
\includegraphics[scale=0.2]{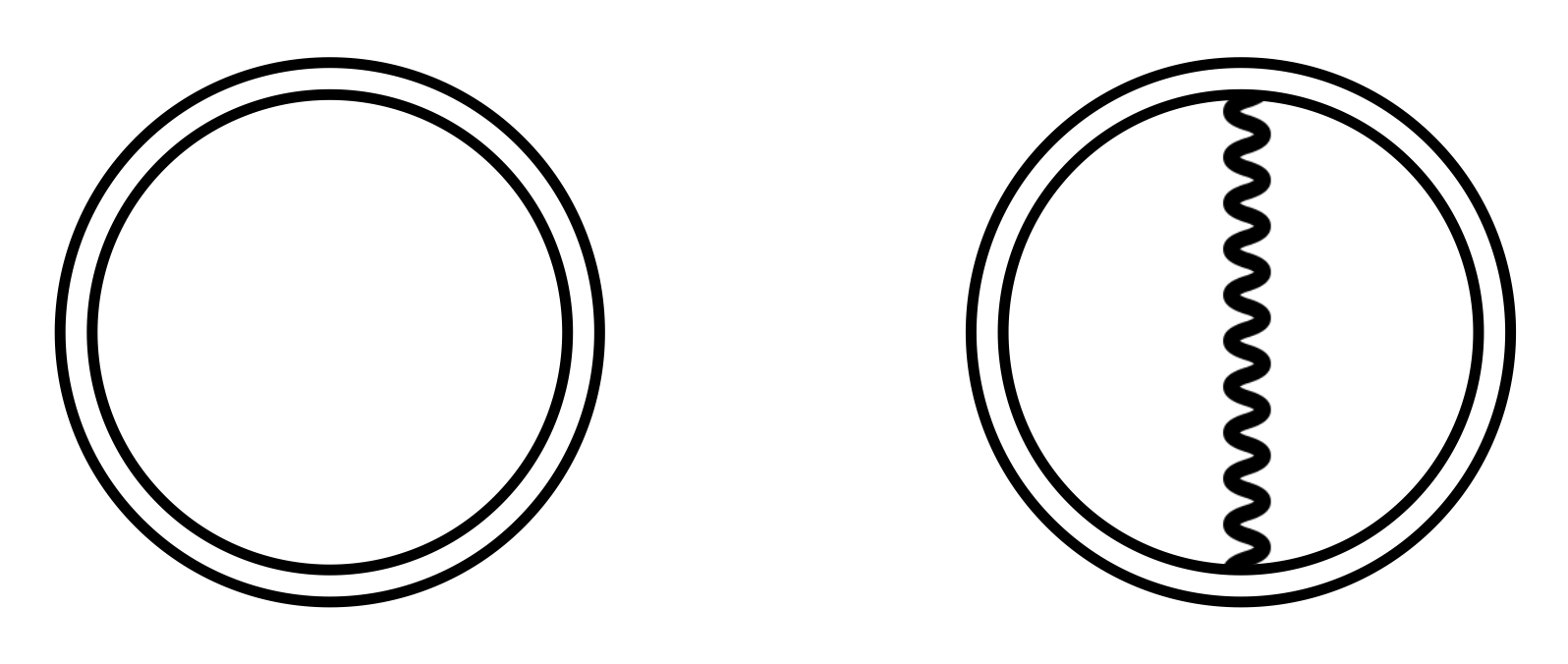}
\caption{The irreducible one-loop (left) and two-loop (right) diagrams contributing to the Euler-Heisenberg effective Lagrangian, with the double solid lines representing the fully-dressed fermion propagator.}
\label{fig:1and2loop}
\end{figure}

The exact renormalized one-loop QED effective Lagrangian in a uniform background electromagnetic field  
was computed long ago by Euler and Heisenberg \cite{euler,weisskopf,schwinger,dittrich-reuter,dunne-kogan}.
This  computation is made possible by the existence of a simple exact integral representation of the electron propagator in a constant background field \cite{fock,nambu}. When the constant background field is purely magnetic,
the one-loop effective Lagrangian is real, but when the constant background field is electric,
the one-loop effective Lagrangian has both a real and imaginary part.
The non-perturbative imaginary part determines the
rate of electron-positron pair-production from vacuum
\cite{euler,schwinger}.
The one-loop Euler-Heisenberg expression in \cite{euler} is an explicit Borel-Laplace integral representation, so the non-perturbative properties can be extracted straightforwardly from the singularity structure of the exact Borel transform function, which has only pole singularities. This meromorphic property of the Borel transform no longer holds at higher loop order.
The two-loop Euler-Heisenberg effective Lagrangian, which includes the new effect of photon exchange between the virtual particles in the fermion loop (see Figure \ref{fig:1and2loop}), was first calculated
by Ritus \cite{ritus1,ritus2,ritus3,lebedev-ritus}, also based on the exact proper-time integral representation of the electron propagator.  A new feature at two-loop order is the necessity of mass renormalization, and Ritus found an exact two-parameter integral expression incorporating both charge and mass renormalization.  See also \cite{fliegner,kors}. 
Ritus's two-loop expression is not explicitly in the form of a Borel-Laplace integral, so the extraction of non-perturbative properties is less direct than at one-loop order. In this paper we discuss the extraction of this non-perturbative information at two-loop, extending the analysis of \cite{ritus1,ritus2,ritus3,lebedev-ritus,ds-2loop,Dunne:2004xk}. The Borel transform is not meromorphic at two-loop, and the non-perturbative structure is different in several interesting ways. 
This distinction continues at all higher loop orders.
Since the constant background field fermion propagator is known exactly, in principle one can express the $l$-loop Euler-Heisenberg effective Lagrangian as a $2(l-1)$-fold parametric integral. However, already at the three-loop level only partial exact results are known for the fully renormalized effective Lagrangian \cite{krasnansky,Huet:2009cy,Huet:2017ydx, Huet:2018ksz}.
In this paper we propose a possible new approach to this problem, based on ideas from resurgent Borel-\'Ecalle asymptotic analysis, which enables the decoding of non-perturbative information directly from perturbative information, not only from an exact integral representation.

We are motivated in part by the general goal of understanding more deeply the structure of the QED perturbative expansion, but also by pragmatic questions concerning the behavior of QED in the ultra-intense limit, which are directly relevant for planned experiments at both DESY \cite{luxe} and SLAC \cite{Yakimenko:2018kih,facet} to probe non-linear and non-perturbative effects arising in interactions involving high-intensity lepton beams and lasers. This new experimental regime promises many surprises, and presents significant theoretical challenges \cite{DiPiazza:2011tq,Meuren:2020nbw}. Deeper understanding of the strong field limit of effective Lagrangians at higher order in the fine structure constant $\alpha$ may shed light on scattering amplitudes in strong background fields, in particular those associated with high intensity lasers. For example, seminal work by Ritus and Narozhnyi has made predictions for the resulting structure at higher loop order for the special case where the background laser field is represented as a constant crossed field \cite{Narozhnyi:1979at,Narozhnyi:1980dc,Morozov:1981pw}. The physical impact of these results is an active area of investigation \cite{Fedotov:2016afw,Podszus:2018hnz,Ilderton:2019kqp,Mironov:2020gbi}, seeking to build on the pioneering analysis of Ritus and Narozhnyi. 

The all-orders Euler-Heisenberg effective Lagrangian can be written as a series in $\alpha$, the fine structure constant\footnote{We use natural units ($\hbar=c=\varepsilon_0=1$), with the fine-structure constant $\alpha=e^2/4\pi$. In keeping with the common convention for perturbative QED expansions, e.g. for the anomalous magnetic moment, we write the perturbative QED expansion parameter as $\alpha/\pi$, and rescale the QED beta function coefficients correspondingly.}
\begin{eqnarray}
{\mathcal L}\left(\alpha, \frac{e F}{m^2}\right) \sim \sum_{l=1}^\infty 
\left(\frac{\alpha}{\pi}\right)^l 
{\mathcal L}^{(l)}\left(\frac{e F}{m^2}\right)
\label{eq:alpha-exp}
\end{eqnarray}
where $F$ denotes the strength of the constant background field, which we consider here to be {\it either} magnetic {\it or} electric.
At a fixed loop order $l$, corresponding to a given order in $\alpha$, the weak field expansion of the effective Lagrangian ${\mathcal L}^{(l)}\left(\frac{e F}{m^2}\right)$ is an asymptotic series with perturbative coefficients $a_n^{(l)}$:
\begin{eqnarray}
{\mathcal L}^{(l)}\left(\frac{e F}{m^2}\right) \sim 
% \left(\frac{\alpha}{\pi}\right)^l 
 \frac{\pi^{2(l-2)}F^2}{(l-1)!}\qty(\frac{e F}{m^2})^2\sum_{n=0}^\infty a_n^{(l)} \qty(\frac{e F}{m^2})^{2n}
\qquad, \quad e F\ll m^2
\label{eq:l-weak}
\end{eqnarray}
Here we have chosen a particular normalization of the expansion coefficients $a_n^{(l)}$, the motivation for which is explained below: see the discussion following (\ref{all-loop-imag}).

Thus, the full expansion (\ref{eq:alpha-exp}) is a double series, an expansion both in $\alpha$ and in the field strength.\footnote{In fact, in general it is a {\it triple} expansion since for each loop order $l$  the weak field expansion (\ref{eq:l-weak}) is itself a double series expansion in the two Lorentz invariant combinations of the constant background field. Here, for simplicity, we concentrate on a constant background field that is either magnetic or electric, but not both.} 
We have separated a factor of $\left(\frac{\alpha}{\pi}\right)^l$ from our definition of  ${\mathcal L}^{(l)}\!\left(\frac{e F}{m^2}\right)$, to emphasize the fact that the loop expansion  (\ref{eq:alpha-exp}) is a perturbative expansion, which is also expected to be asymptotic \cite{dyson}.
If the background field is magnetic, of magnitude $B$, at $l$-loop order ${\mathcal L}^{(l)}\!\left(\frac{e B}{m^2}\right)$ is expected to be unambiguously Borel summable to a real expression. If the background field is electric, of magnitude $E$, at $l$-loop order ${\mathcal L}^{(l)}\!\left(\frac{e E}{m^2}\right)$ is expected to have both real and imaginary parts, also expressible as a well-defined Borel representation. Furthermore, it is expected that at each loop order the electric background field result can be obtained by analytic continuation $B\to iE$ from the real magnetic field result.\footnote{For a constant field of {\it either} magnetic {\it or} electric nature, the relevant Lorentz invariant quantity is $E^2-B^2$.} 
Here we analyze these issues beyond the familiar one-loop result, using ideas and methods from Borel-\'Ecalle summation \cite{Ecalle,costin-book,Marino:2012zq,Sauzin,gokce,Costin:2019xql,Costin:2020hwg,Costin:2020pcj}.

At one-loop order  the non-perturbative imaginary part has 
a well-known polylogarithm expression which can be expanded as a {\it convergent} weak-field instanton expansion:
\begin{eqnarray}
\label{eq:l1-imag}
{\rm Im}\left[ \mathcal{L}^{(1)}\left(\frac{e E}{m^2}\right)\right]&=&\frac{E^2}{2\pi}\, \text{Li}_2\qty(e^{-\pi m^2/(eE)})
\\
&=&
\frac{E^2}{2\pi}\left\{ e^{- \pi m^2/(eE)} +\frac{1}{2^2} e^{-2\pi m^2/(eE)}+\frac{1}{3^2} e^{-3\pi m^2/(eE)} +\ldots\right\}
\label{eq:l1-imag1}
\end{eqnarray}
Note that the instanton terms are not multiplied by fluctuation series, just numerical residue factors, so this is a very simple example of a trans-series.
By contrast, at two-loop order the weak-field instanton expansion of the non-perturbative imaginary part is conjectured to be of the form \cite{ritus1,ritus2,ritus3,lebedev-ritus}
\begin{eqnarray}
{\rm Im}\left[{\mathcal L}^{(2)}\left(\frac{e E}{m^2}\right)\right] \sim
\frac{\pi E^2}{2}\,  \left\{ e^{-\pi m^2/(eE)}\bigg(1+\dots\bigg) +
\sqrt{\frac{m^2}{eE}}\sum_{k=2}^\infty e^{-k\pi m^2/(eE)}\left(-c_k+\sqrt{\frac{e E}{m^2}}+\ldots\right)\right\}
\label{eq:l2-imaginary}
\end{eqnarray}
where the numerical coefficient $c_k$, for $k\geq 2$, is
\begin{eqnarray}
c_k=\frac{1}{2\sqrt{k}}\sum_{\ell=1}^{k-1}\frac{1}{\sqrt{\ell(k-\ell)}} \qquad, \quad k\geq 2
\label{eq:ritus-cn}
\end{eqnarray}
This conjectured two-loop structure is quite different from one-loop: there are (unspecified) fluctuations about the one-instanton term, and at higher instanton orders ($k\geq 2$) there is a stronger weak-field prefactor, $\sqrt{\frac{m^2}{eE}}$, followed also by fluctuations (only partially specified). Our goal here is to investigate these  fluctuation terms in the non-perturbative imaginary part for the two-loop effective Lagrangian in an electric background field, by analytic continuation from the weak field expansion for a magnetic background. We focus on reconstructing non-perturbative information from {\it finite-order} perturbative information, rather than the standard method of seeking an exact closed-form (multi-)integral representation, which currently appears prohibitively difficult even at three-loop order \cite{Huet:2009cy,Huet:2017ydx, Huet:2018ksz}.  The motivation for this different approach is as a proof-of-principle test for a new approach to Euler-Heisenberg computations at 
higher loop order. 

Note that while $e F$ is a renormalization group invariant, both $\alpha$ and $m^2$ are scale dependent. At one-loop order, $m^2$ is simply the bare mass and the charge renormalization is conventionally done at the physical electron mass  scale. At two-loop order, a field-dependent mass renormalization is required, and ${\mathcal L}^{(2)}$ is again conventionally renormalized at the physical electron mass scale \cite{ritus1,ritus2,ritus3}. Being a doubly-perturbative expansion, it is not immediately obvious how to extract strong-field or short-distance or non-perturbative information from the full all-orders result (\ref{eq:alpha-exp}). Different expansions, for example at fixed order in $\alpha$ or in $e F$, are possible, and correlated or uniform limits may also be physically relevant in certain circumstances.
For example, 
 if we sum the {\it leading} weak field contributions to the imaginary part of ${\mathcal L}^{(l)}\left(\frac{e E}{m^2}\right)$ for an electric background, this sum exponentiates to $e^{\alpha \pi}$ times the leading one-loop result \cite{ritus3,Affleck:1981bma,Huet:2017ydx}:
 \begin{eqnarray}
{\rm Im}\left[ \mathcal{L}\left(\alpha, \frac{e E}{m^2}\right)\right]\sim
 \frac{\alpha E^2}{2\pi^2}\, e^{\alpha \pi}\, e^{- \pi m^2/(eE)} +\dots
\qquad, \quad e E\ll m^2
\label{all-loop-imag}
\end{eqnarray}
This exponential factor can be computed from the world-line representation of the effective Lagrangian \cite{Affleck:1981bma}, or can be understood \cite{ritus3} as encoding the leading field-dependent mass shift at two-loop order: $m^2\to m^2-\alpha/(e E)$. By a Borel dispersion relation, this exponentiation can be translated into a conjecture for the leading large-order growth of the perturbative expansion coefficients $a_n^{(l)}$ \cite{Dunne:2004xk}, which motivates the choice of overall normalization of these coefficients in (\ref{eq:l-weak}).

Here we propose a different approach to the higher-loop computations, based on expressing the perturbative weak magnetic field expansion at a given loop order as an {\it approximate} Borel integral, whose strong-field limit can be extracted with surprisingly high precision, and whose analytic continuation from a magnetic to an electric background can also be achieved with high precision, including exponentially suppressed non-perturbative information. Such an approach relies on efficient and near-optimal methods to perform the necessary analytic continuations  \cite{Costin:2019xql,Costin:2020hwg,Costin:2020pcj}.
In this paper we test the feasibility of such methods applied to the Euler-Heisenberg effective Lagrangian at one-loop and two-loop, and we conclude with comments about the prospects for higher loop orders. 

\section{The Exact One-loop Euler-Heisenberg Effective Lagrangian}

\subsection{Exact Results at One-Loop Order}

We first review well-known properties of the one-loop  Euler-Heisenberg QED effective Lagrangian in a constant background magnetic field, $B$,  conventionally expressed as a proper-time integral \cite{euler,dunne-kogan}
\begin{eqnarray}
\mathcal{L}^{(1)}\left(\frac{e B}{m^2}\right)=- \frac{B^2}{2}\int_0^\infty \frac{\dd{t}}{t^2}\qty(\coth t-\frac{1}{t}-\frac{t}{3})e^{-m^2 t/(eB)}
\label{eq:l1}
\end{eqnarray}
[Recall our notational convention that a factor of $\left(\frac{\alpha}{\pi}\right)^l$ is extracted at $l$-loop order]. The weak field expansion of (\ref{eq:l1}) is a prototypical effective field theory expansion, expressing the physics of the light fields (photons) after integrating out the heavy fields (electrons/positrons) at the electron mass scale $m$:
\begin{eqnarray}
\mathcal{L}^{(1)}\left(\frac{e B}{m^2}\right) \sim  \frac{B^2}{\pi^2}\qty(\frac{eB}{m^2})^2\sum_{n=0}^\infty a_n^{(1)} \qty(\frac{eB}{m^2})^{2n}
\qquad, \quad eB\ll m^2
\label{eq:l1-weak}
\end{eqnarray}
Here the factorially divergent one-loop expansion coefficients are known exactly:
\begin{eqnarray}
a_n^{(1)}&=& (-1)^n \frac{\Gamma(2n+2)}{\pi^{2n+2}}\zeta(2n+4)
\label{eq:a1}\\
&=& (-1)^n \frac{\Gamma(2n+2)}{\pi^{2n+2}}\left(1+\frac{1}{2^2}\cdot\frac{1}{2^{2n+2}}+\frac{1}{3^2}\cdot\frac{1}{3^{2n+2}}+ \frac{1}{4^2}\cdot\frac{1}{4^{2n+2}}+\dots\right) 
\label{eq:a12}
\end{eqnarray}
The corrections to the leading factorial growth  in (\ref{eq:a12}) 
are {\it exponential} in $n$,  which is directly related to the non-appearance of power-law fluctuation corrections in the instanton expansion  (\ref{eq:l1-imag1}) \cite{dunne-kogan}. We have deliberately written the Riemann zeta factor in the form in (\ref{eq:a12})  to emphasize the correspondence between the $\frac{1}{k^2}$ factors multiplying the $k$-instanton terms in (\ref{eq:l1-imag1}), and the residues of the singularities of the Borel transform. See also Figure \ref{fig:b1-imaginary}.

The exact integral representation (\ref{eq:l1}) can also be expanded in the strong field limit, yielding a convergent strong field expansion whose leading behavior is:
\begin{eqnarray}
\mathcal{L}^{(1)}\left(\frac{e B}{m^2}\right) \sim
\frac{1}{3} \cdot  \frac{B^2}{2}\qty(\ln\left(\frac{eB}{\pi m^2}\right) -\gamma+\frac{6}{\pi^2}\zeta'(2))
\quad, \quad e B\gg m^2
\label{eq:l1-strong}
\end{eqnarray}
where $\gamma\approx 0.5772...$ is the Euler-Mascheroni constant, and $\zeta'(2)=\frac{1}{6} \pi ^2 (-12 \log (A)+\gamma +\ln (2\pi ))\approx -0.937548$, with $A$ being
the Glaisher-Kinkelin constant. 
The coefficient $\frac{1}{3}$ of the logarithmic term in (\ref{eq:l1-strong}) is the one-loop QED beta function coefficient $\beta_1$, associated with one-loop charge renormalization \cite{weisskopf,dunne-kogan}: 
\begin{eqnarray}
\beta_{\rm QED}(\alpha) =2\alpha\sum_{n=1}^\infty \beta_n \left(\frac{\alpha}{\pi}\right)^n =2\alpha\qty[\frac{1}{3}\left(\frac{\alpha}{\pi}\right) +\frac{1}{4}\left(\frac{\alpha}{\pi}\right)^2 +\dots] 
\label{eq:beta}
\end{eqnarray}

The weak field expansion (\ref{eq:l1-weak}) is an asymptotic series \cite{schwinger,ioffe,ogievetsky,graffi,chadha,dunne-kogan}, whose Borel sum is the one-loop Euler-Heisenberg integral representation (\ref{eq:l1}), with one-loop Borel transform function:
\begin{eqnarray}
{\mathcal B}^{(1)}(t): =-\frac{1}{\pi^2\, t^2}\qty(\coth(\pi t)-\frac{1}{\pi t}-\frac{\pi t}{3})
&=&\frac{2}{\pi^3} \sum_{n=0}^\infty (-1)^n \zeta(2n+4)\, t^{2n+1}
\label{eq:borel1a}\\
&=& \frac{2}{\pi^3}  \sum_{k=1}^\infty \frac{t}{k^2(t^2+k^2)}
\label{eq:borel1b}
\end{eqnarray}
Here we have chosen to make the convenient rescaling of the Borel variable by a factor of $\pi$, to absorb the powers of $1/\pi$ in (\ref{eq:a1})-(\ref{eq:a12}), which has the effect of placing the Borel poles at integer multiples of $i$, rather than at integer multiples of $ \pi i$. Then the exact one-loop effective Lagrangian is recovered via the Borel-Laplace integral (note the extra factor of $\pi$ in the exponent):
\begin{eqnarray}
\mathcal{L}^{(1)}\left(\frac{e B}{m^2}\right)=\frac{\pi B^2}{2}\int_0^\infty dt\, e^{-m^2 \pi t/(eB)} \, {\mathcal B}^{(1)}(t)
\label{eq:l1pi}
\end{eqnarray}
The small $t$ expansion (\ref{eq:borel1a}) of ${\mathcal B}^{(1)}(t)$  generates the asymptotic weak magnetic field expansion (\ref{eq:l1-weak}), while the partial-fraction expansion in (\ref{eq:borel1b}) exhibits the meromorphic nature of the one-loop Borel transform function ${\mathcal B}^{(1)}(t)$, with 
an infinite line of integer-spaced simple pole singularities along the imaginary Borel axis, at $t=\pm i k$, for $k=1, 2, 3, \dots$. Therefore, under analytic continuation, $B\to i E$, where $E$ is a constant background electric field,  these poles lead to the non-perturbative imaginary part of the effective Lagrangian in (\ref{eq:l1-imag}), with weak-field expansion in (\ref{eq:l1-imag1}). In the strong electric field limit the leading behavior can be obtained by analytic continuation from the strong magnetic field expansion in (\ref{eq:l1-strong}):
\begin{eqnarray}
{\rm Im}\left[ \mathcal{L}^{(1)}\left(\frac{e E}{m^2}\right)\right]
\sim
\beta_1 \left(\frac{\pi}{2}\right) \frac{E^2}{2} \quad,  \quad e E\gg m^2 
\label{eq:l1-imaginary}
\end{eqnarray}
consistent with the strong field limit of the exact polylog expression in (\ref{eq:l1-imag}). 
This also follows from the explicit representation of the one-loop effective Lagrangian in terms of the Barnes gamma function $G$ \cite{dunne-kogan}:
\begin{align}
\mathcal{L}^{(1)}\qty(\frac{eB}{m^2})&=\frac{2B^2}{m^4}\left[-\frac{1}{12}+\zeta'(-1)+\frac{1}{16}\qty(\frac{m^2}{eB})^2+\qty(-\frac{1}{12}+\frac{m^2}{4eB}-\frac{1}{8}\qty(\frac{m^2}{eB})^{2})\ln\frac{m^2}{2eB}\right.\nonumber\\
&\quad \Bigg.-\qty(1-\frac{m^2}{2eB})\ln\Gamma\qty(\frac{m^2}{2eB})-\ln G\qty(\frac{m^2}{2eB})\Bigg]
\label{eq:barnes}
\end{align}
This Barnes representation of the one-loop effective Lagrangian is particularly convenient for analytic continuation of $B$, since the analytic properties of the Barnes G function are well known \cite{barnes,nemes}.

\subsection{Borel Analysis at One-Loop Order}
\label{sec:1l-borel}

In preparation for the two-loop analysis, where a simple Borel representation of the form in (\ref{eq:l1}) is not available, and a closed-form expression such as (\ref{eq:barnes}) in terms of a special function like the Barnes function is not known, we ask how we can recover accurate {\it approximations} to the various exact results listed in the previous sub-section. We do this here at one-loop and then extend these methods to two-loop order in Section \ref{sec:twoloop}.
Specifically, we begin with just a {\it finite number} of terms of the one-loop perturbative weak magnetic field expansion in (\ref{eq:l1-weak}), and seek to recover:
\begin{enumerate}
\item
the strong magnetic field limit in (\ref{eq:l1-strong}) [this is a weak-field to strong-field extrapolation];
\item
the non-perturbative imaginary part  (\ref{eq:l1-imag1}) of the effective Lagrangian in an electric field background [this is analogous to a Euclidean to Minkowski analytic continuation].
\end{enumerate}
We develop a modified Pad\'e-Borel approach that leads to remarkable precision with surprisingly little input information. Note that \cite{florio} has already showed that Pad\'e-Borel summation achieves very accurate extrapolations of the weak magnetic field series for the one-loop Euler-Heisenberg effective Lagrangian. However, for the analysis at two-loop we require even higher precision; hence the new procedure described below.

\begin{figure}[htb!]
\centering
\includegraphics[scale=.4]{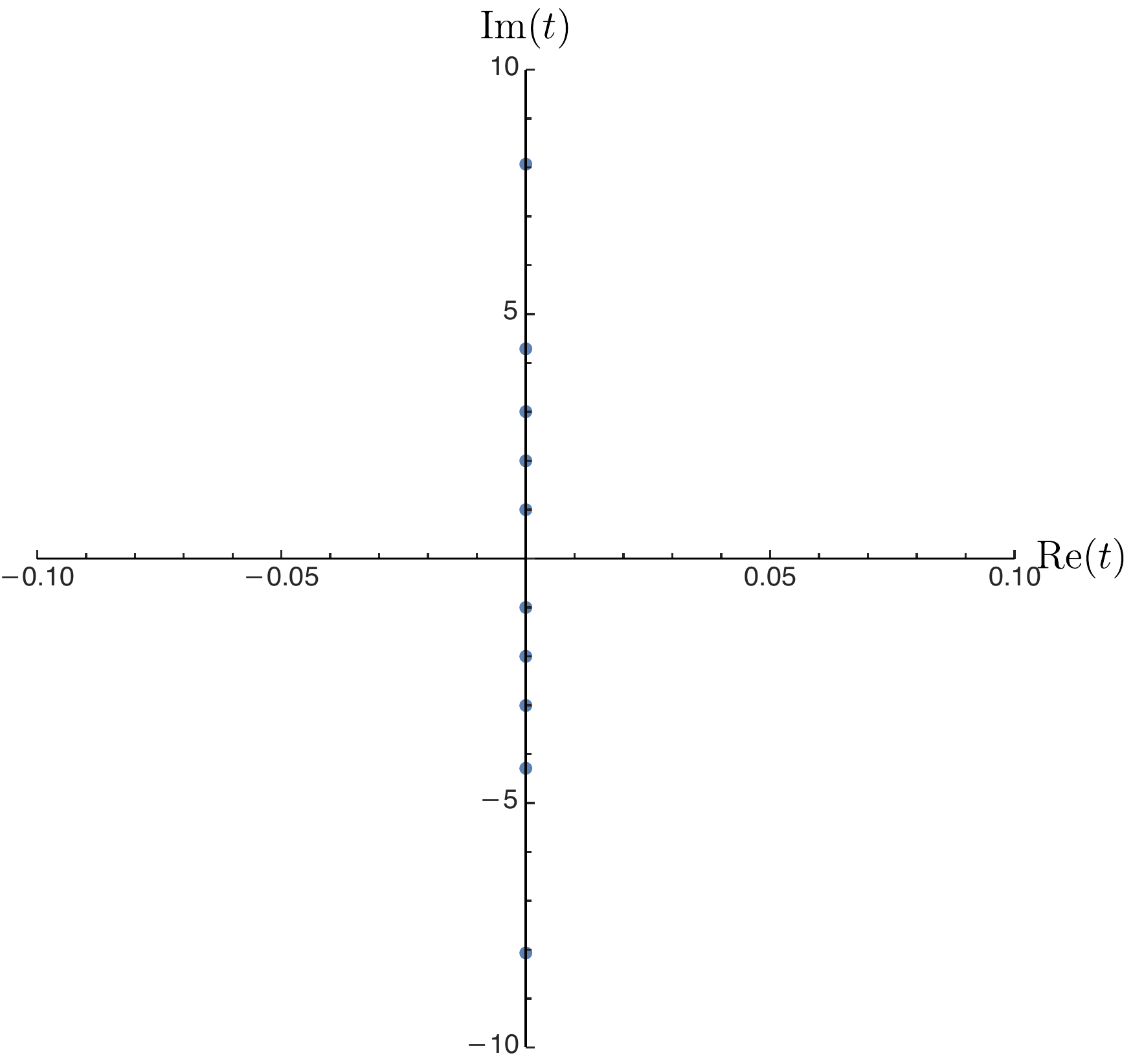}
\includegraphics[scale=.4]{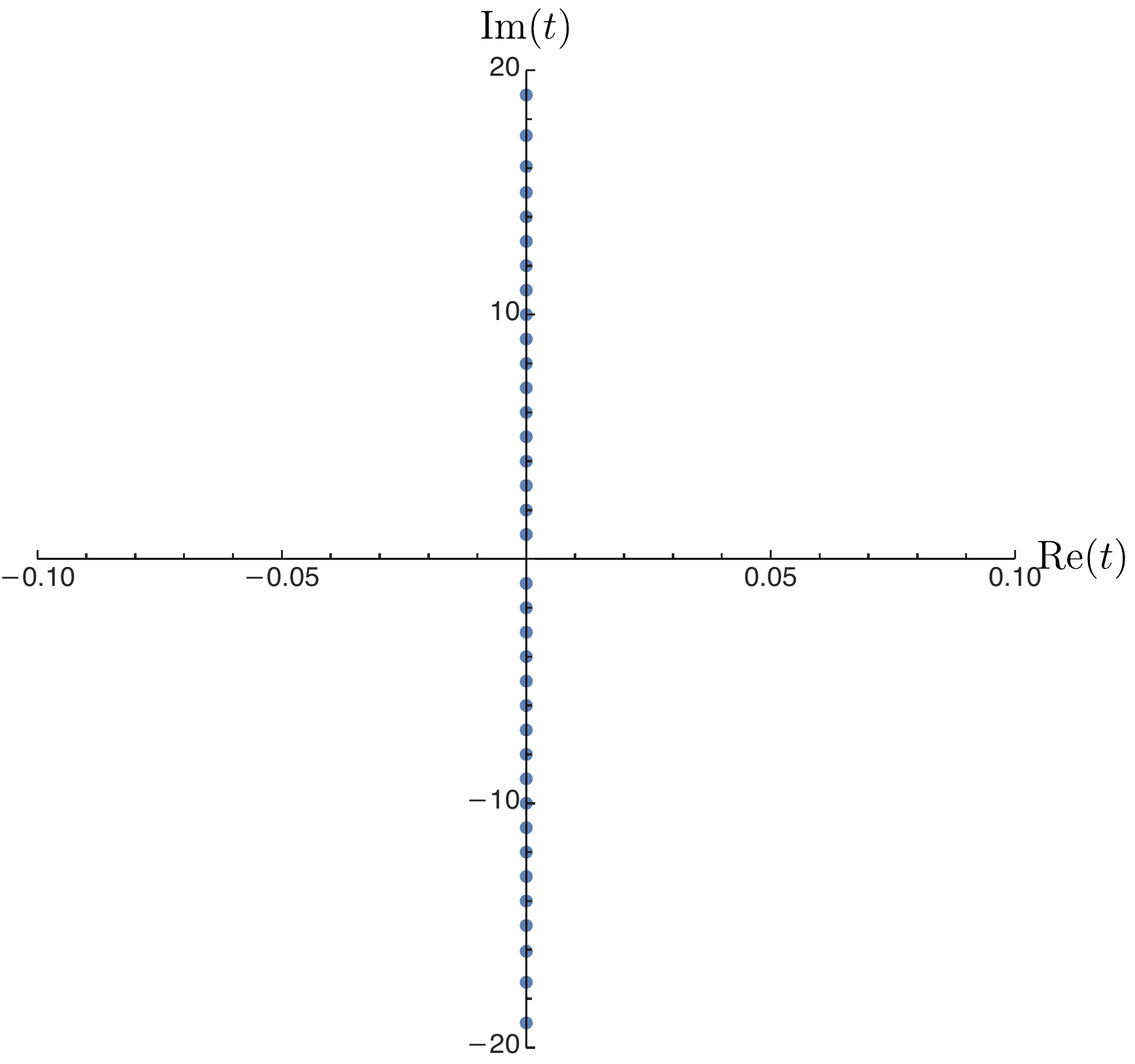}
\caption{Poles of the Pad\'e approximation $\mathcal{PB}_N^{(1)}(t)$ in (\ref{eq:pade-borel1}) for the truncated Borel transform, shown for both $N=10$ (left) and $N=50$ (right). Note that all the poles lie on the imaginary axis and tend towards integer multiples of $\pm i$. This can be contrasted with two-loop case in Figure \ref{fig:PB2-poles}.}
\label{fig:PB1-poles}
\end{figure}

Given only a finite number of terms of the weak magnetic field expansion (\ref{eq:l1-weak}), or equivalently only a finite number of terms in the small $t$ expansion of the Borel transform in (\ref{eq:borel1a})-(\ref{eq:l1pi}), the key to an accurate extrapolation to other regions of the complex $B^2$ plane (recall that $B^2<0$ corresponds to an electric field background) is to have an accurate analytic continuation of the truncated Borel transform function
\begin{eqnarray}
\mathcal{B}^{(1)}_N(t):=\frac{2}{\pi^2}\sum_{n=0}^{N-1} \frac{a_n^{(1)}}{(2n+1)!}\qty(\pi t)^{2n+1}
\label{eq:borel1-n}
\end{eqnarray}
Since the one-loop Borel transform is meromorphic, the optimal approach \cite{Costin:2020hwg,Costin:2020pcj} is to use a Pad\'e approximation,\footnote{While a Pad\'e approximation could be applied directly to the truncated asymptotic weak magnetic field expansion (\ref{eq:l1-weak}), a significantly better extrapolation \cite{Costin:2020hwg,Costin:2020pcj} is achieved by the Pad\'e-Borel method \cite{zinn-book}, making a Pad\'e approximation in the Borel $t$ plane rather than a Pad\'e approximation in the original physical variable $eB/m^2$.} which expresses $\mathcal{B}^{(1)}_N(t)$ as a ratio of polynomials:
\begin{eqnarray}
\mathcal{P}_{[L, M]}\left({\mathcal B}^{(1)}_N(t)\right) =\frac{P^{(1)}_{L}(t)}{Q^{(1)}_{M}(t)} \quad, \quad \text{where}\quad \frac{P^{(1)}_{L}(t)}{Q^{(1)}_{M}(t)}=\mathcal{B}^{(1)}_N(t)+\mathcal{O}\left(t^{L+M+1}\right)
\label{eq:pade}
\end{eqnarray}
where $L+M=2N+1$.
The zeros of the denominator polynomial $Q_{M}(t)$ approximate the true singularities (see (\ref{eq:borel1b})) of $\mathcal{B}^{(1)}(t)$, which lie at $t=\pm ik$ for $k\neq 0\in\mathbbm{Z}$ due to our rescaling $t\rightarrow \pi t$. See Figure \ref{fig:PB1-poles}. Tables of the one-loop Pad\'e-Borel poles for $N=10$ input terms and for $N=50$ input terms are shown below in (\ref{eq:PB1-pole-table10}) and (\ref{eq:PB1-pole-table50}), in which we see poles stabilizing along the imaginary Borel axis at integer multiples of $\pm i$.

Pad\'e poles from 10 input terms:
\begin{gather}
\pm\left\{ 1.0000 i,  2.0000 i,  3.0057 i,
4.2905 i, 8.0671 i\right\}
   \label{eq:PB1-pole-table10}
   \end{gather}

Pad\'e poles from 50 input terms:
\begin{align}
&\pm\left\{ 1.0000 i, 2.0000 i, 3.0000 i, 4.0000 i, 5.0000 i, 6.0000 i, 7.0000 i, 8.0000 i,   \right.\nonumber\\
& \left.\quad\ 9.0000 i, 10.000 i, 11.000 i, 12.000 i, 13.000 i, 14.000 i, 15.005 i, 16.062 i, 17.326 i,  \right.\label{eq:PB1-pole-table50}\\
& \left.\quad\ 18.990 i,  21.239 i, 24.345 i, 28.798 i, 35.592 i, 47.048 i, 70.145 i, 139.79 i\right\}\nonumber
\end{align}

We dramatically improve the quality of the extrapolation if we take advantage of known information about the opposite (strong field) limit, which corresponds to including information about the $t\to \infty$ behavior of $\mathcal{B}^{(1)}(t)$. Here we appeal to the fundamental physical interpretation of the logarithmic behavior of the strong field expansion (\ref{eq:l1-strong}) in terms of charge renormalization and the conformal anomaly \cite{euler,weisskopf,ritus1,ritus2,dunne-kogan}, which translates into the requirement that
\begin{figure}[b]
\centering
\includegraphics[scale=.7]{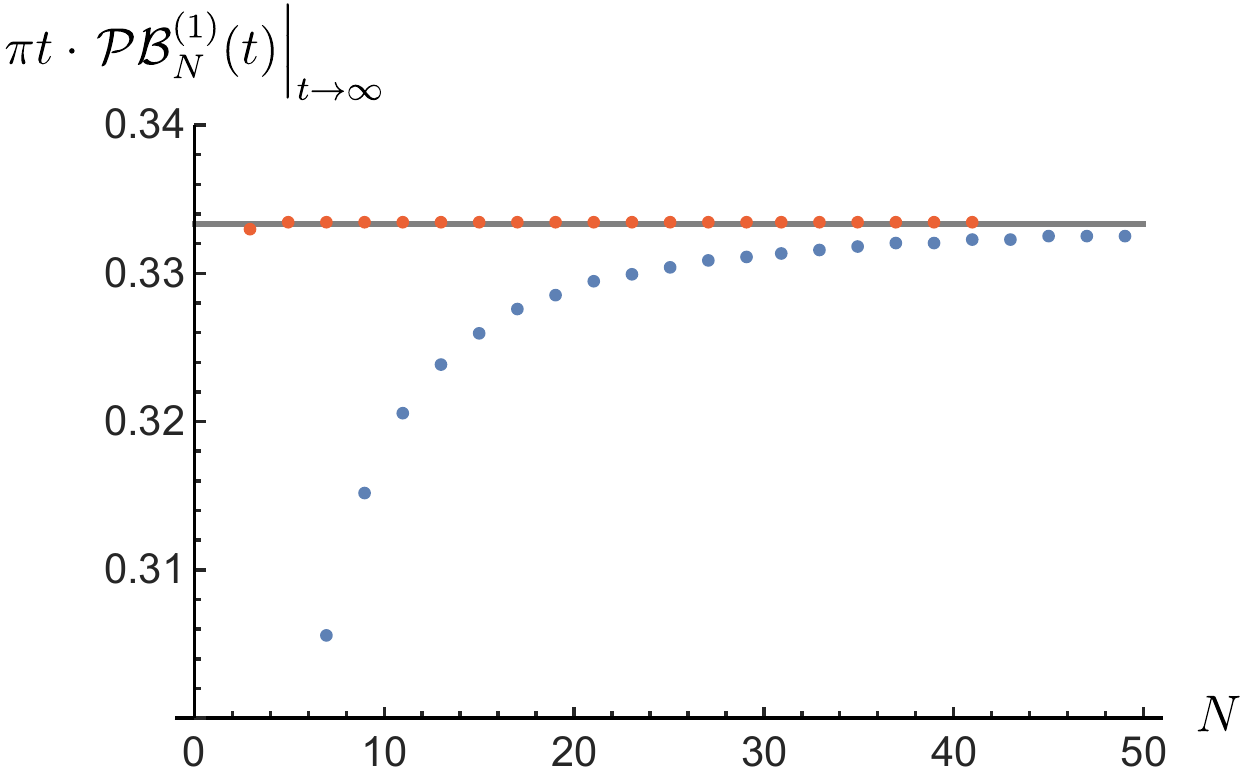}
\caption{$N$ dependence of the limiting value $\lim\left[\pi t\cdot \mathcal{PB}_N^{(1)}(t)\right]_{t\to\infty}$, for $N$ ranging from 1 to 50, for the near-diagonal Pad\'e approximant of the truncated one-loop Borel transform function in (\ref{eq:pade-borel1}).  The blue dots indicate the original values obtained from expanding $\pi t\cdot\mathcal{PB}_N^{(1)}(t)$ about $t\rightarrow\infty$, and which appear to be tending towards $1/3$. The red dots show a 4-th order Richardson extrapolation of this data. Observe that the physical value $\beta_1=1/3$ is approached quite closely already for  $N\approx 10$.}
\label{fig:beta1}
\end{figure}
\begin{eqnarray}
\mathcal{B}^{(1)}(t) \sim \frac{\beta_1}{\pi t} +\dots \quad, \quad t\to+\infty
\label{eq:borel1-large-t}
\end{eqnarray}
We therefore choose an off-diagonal Pad\'e-Borel approximant with $M=L+1=N+1$
\begin{eqnarray}
\mathcal{PB}^{(1)}_{N}(t):=\frac{P^{(1)}_{N}(t)}{Q^{(1)}_{N+1}(t)}
\label{eq:pade-borel1}
\end{eqnarray}
Note that we do not impose that the overall coefficient of $\pi t\cdot \mathcal{PB}^{(1)}_{N}(t)$ in the limit $t\to+\infty$ be equal to the physical value, $\beta_1=\frac{1}{3}$. We simply impose the functional form that $\pi t\cdot \mathcal{PB}^{(1)}_{N}(t)\to \text{constant}$, as $t\to +\infty$. Remarkably, the physical value, $\beta_1$, emerges in the large $N$ limit, already at $N\approx 10$.  See Figure \ref{fig:beta1}.

\begin{figure}[tb!]
\centering
\includegraphics[scale=.7]{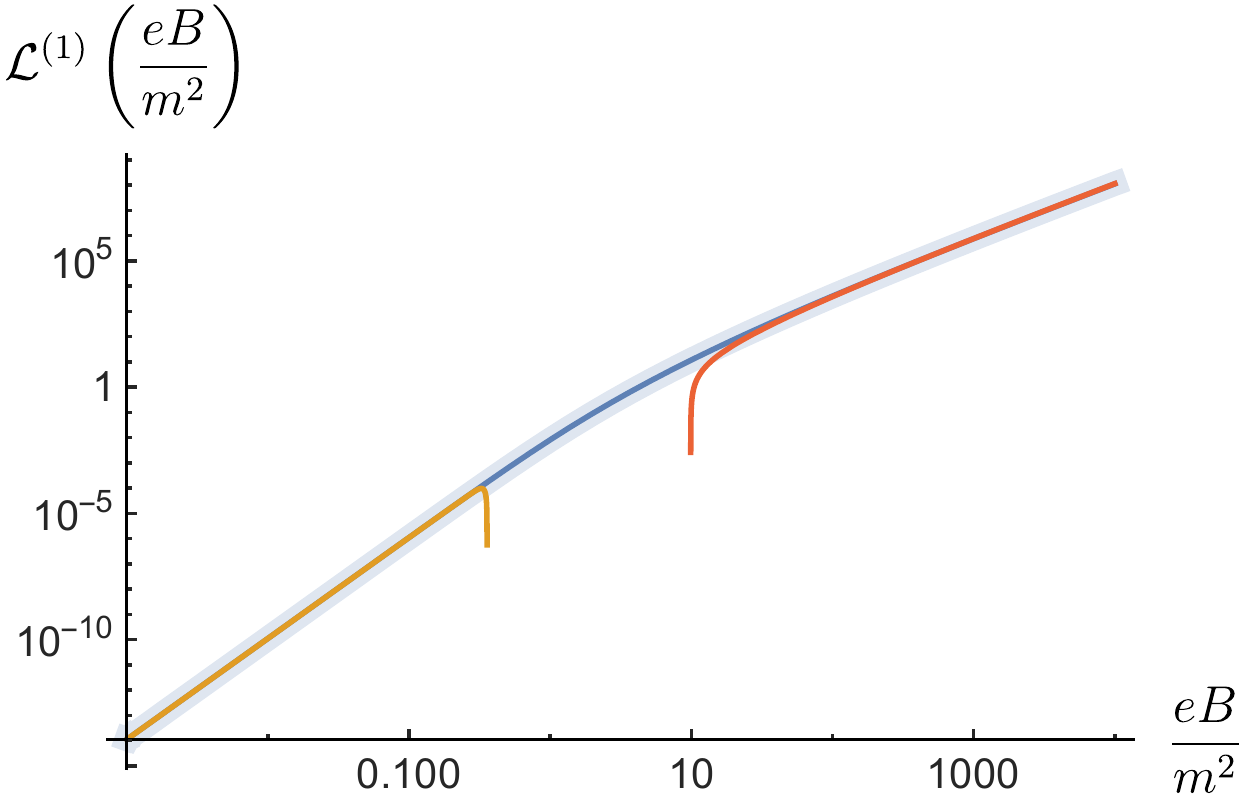}
\caption{The blue curve is a log-log plot of the modified Pad\'e-Borel sum of the truncated weak field expansion in (\ref{eq:pb1}), $\mathcal{L}_N^{(1)}$, plotted here using only $N=10$ perturbative input terms. This is indistinguishable from the exact closed-form Barnes function expression in (\ref{eq:barnes}), plotted here as the translucent blue band. The gold curve shows the weak field expansion (\ref{eq:l1-weak}), truncated at $N=10$. 
The red curve shows the leading strong field behavior of $\mathcal{L}^{(1)}$ in (\ref{eq:l1-strong}). Note that the truncated weak-field expansion fails even below the Schwinger limit $eB\approx m^2$, while the modified Pad\'e-Borel sum
accurately interpolates over many orders of magnitude between the weak-field and strong-field behavior. This plot was made using units in which $e=m^2=1$.\vspace{-1cm}}
\label{fig:magnetic1-interpolation}
\end{figure}

\begin{figure}[b]
\centering
\includegraphics[scale=.7]{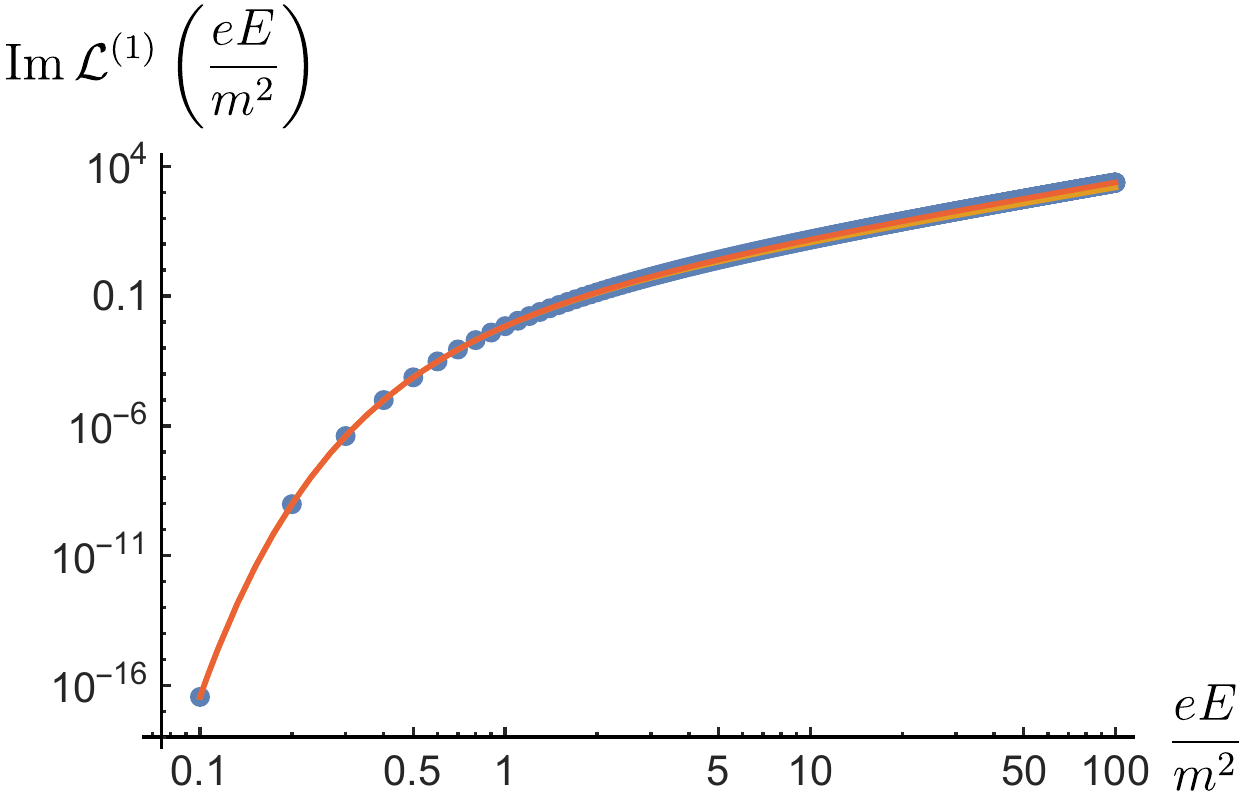}
\caption{A log-log plot of the imaginary part of the electric field effective Lagrangian at one-loop, calculated using $N=10$ (blue dots). The gold curve shows the leading one-instanton contribution, $\propto \exp\qty(-\pi m^2/(eE))$, which displays a small but noticeable deviation in the strong field limit. The red curve is the exact expression for the imaginary part in (\ref{eq:l1-imag}), including the sum over all instantons.  See also Figure \ref{fig:electric1-subleading}. This plot was made using units in which $e=m^2=1$.}
\label{fig:electric1}
\end{figure}

With this procedure, our Pad\'e analytic continuation (\ref{eq:pade-borel1}) of the truncated weak-field expansion (\ref{eq:borel1-n}) leads to an approximate Borel-Laplace integral representation for the one-loop effective Lagrangian as in 
 (\ref{eq:l1pi}):
\begin{eqnarray}
\mathcal{L}_N^{(1)} \left(\frac{e B}{m^2}\right)  =\frac{\pi B^2}{2}\int_0^\infty\dd{t}\, e^{-m^2 \pi t/(e B)}\,  \mathcal{PB}^{(1)}_N\qty(t)
\label{eq:pb1}
\end{eqnarray}
Figure \ref{fig:magnetic1-interpolation} shows the extrapolation of this expression from the weak-field limit to the strong-field limit. Starting with just ten input coefficients of the weak field expansion, the modified Pad\'e-Borel expression in (\ref{eq:pb1}) extrapolates accurately over more than 8 orders of magnitude. This is a significantly farther-reaching extrapolation than in \cite{florio}, due to our improved Pad\'e-Borel transform in (\ref{eq:pade-borel1}). This modified Pad\'e-Borel transform also explains why the pole structure shown in Figure \ref{fig:PB1-poles}, and in Equations (\ref{eq:PB1-pole-table10})-(\ref{eq:PB1-pole-table50}), is much more accurate than that in \cite{florio}, where the physical form of the large $t$ behavior (\ref{eq:borel1-large-t}) was not imposed on the Pad\'e-Borel transform.

Similarly we can use the approximate Borel representation in (\ref{eq:pb1}) to achieve our second goal: analytically continuing from a magnetic background to an electric background, to extract the exponentially small non-perturbative imaginary part of $\mathcal L^{(1)}$ in (\ref{eq:l1-imag}). Since the Borel singularities are all on the imaginary axis, we rotate the Borel contour to 
generate the imaginary part of the one-loop effective Lagrangian for a constant background electric field. Figure \ref{fig:electric1} shows (as blue dots) the result of this calculation, starting with 10 terms of the weak magnetic field expansion. The red curve shows the exact result in (\ref{eq:l1-imag}), summed over all instanton orders, while the gold curve shows the leading one-instanton term. The agreement is excellent in the weak-field limit, and also extrapolates accurately to much stronger fields.

\begin{figure}[tb!]
\centering
\includegraphics[scale=.7]{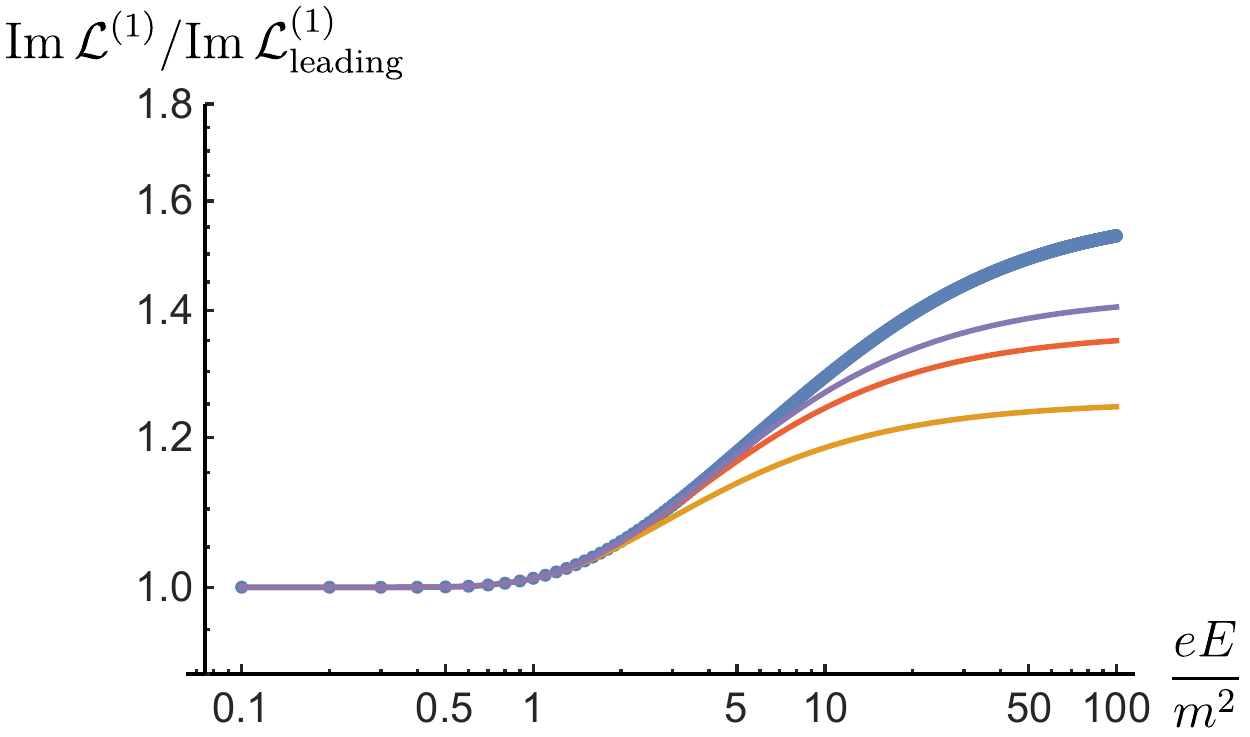}
\caption{The ratio [blue dots] of the imaginary part of the electric field effective Lagrangian $\Im\mathcal{L}^{(1)}\qty(E)$, divided by the leading exponential term from (\ref{eq:l1-imag1}), derived from our extrapolation using as input just $N=10$ terms from the perturbative expression for a magnetic field background. The gold, red, and purple curves show a fit for this ratio, based on one, two, and three exponentially small correction terms, respectively. The fit coefficients are given in (\ref{eq:1loop-fit}), from which it is clear that these sub-leading exponential corrections tend to the known form (\ref{eq:l1-imag}) of the instanton sum. This plot was made using units in which $e=m^2=1$.}
\label{fig:electric1-subleading}
\end{figure}

\begin{figure}[b]
\centering
\includegraphics[scale=.7]{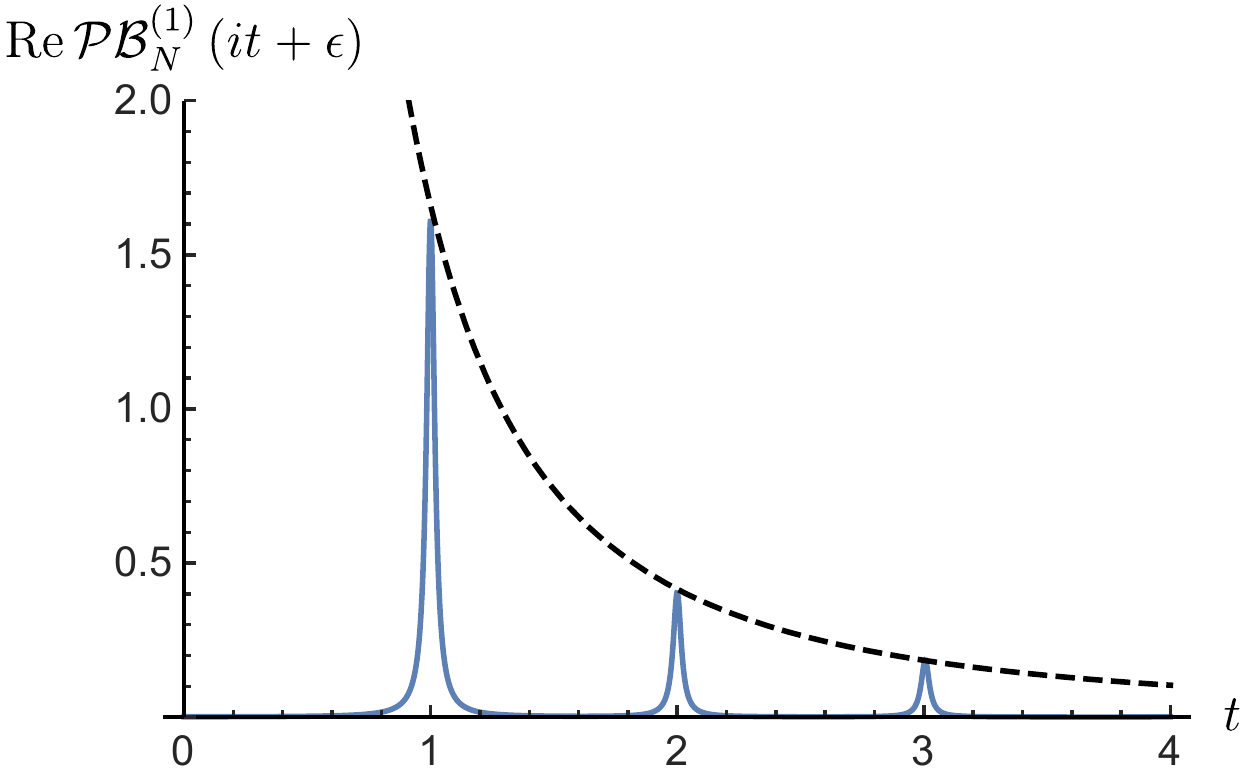}
\caption{Singularity structure of the Borel transform $\mathcal{PB}_N^{(1)}(t)$, for $N=10$, just offset from the imaginary axis, $t\to it+\epsilon$. The plot shows the real part (blue solid curve)
 of $\mathcal{PB}_N^{(1)}(it+ 1/50)$, indicating integer-spaced poles at $t=ik$ along the imaginary Borel axis, with residues falling off quadratically as $1/k^2$ (dashed black curve).}
\label{fig:b1-imaginary}
\end{figure}

In Figure \ref{fig:electric1-subleading} we show that the precision of our extrapolation is sufficiently high that we can probe the {\it exponentially small} higher-instanton corrections to $\Im\mathcal{L}^{(1)}\qty(\frac{eE}{m^2})$, by dividing out 
the one-instanton factor, $E^2/(2\pi) \exp\qty(-\pi m^2/(eE))$. Fitting this ratio with one-term, two-term and three-term exponential fits, we obtain the successively improving approximations:
\begin{align}
&\Im \mathcal{L}^{(1)} \left(\frac{e E}{m^2}\right) \approx \frac{E^2}{2\pi}\left(e^{-\pi m^2/(eE)}+0.253303\, e^{-2\pi m^2/(eE)} +\dots \right) \nonumber \\
&\Im \mathcal{L}^{(1)} \left(\frac{e E}{m^2}\right) \approx \frac{E^2}{2\pi}\left(e^{-\pi m^2/(eE)}+0.249962\, e^{-2\pi m^2/(eE)}+0.114532\, e^{-3\pi m^2/(eE)} +\dots \right)
\label{eq:1loop-fit}\\
&\Im \mathcal{L}^{(1)} \left(\frac{e E}{m^2}\right) \approx \frac{E^2}{2\pi}\left(e^{-\pi m^2/(eE)}+0.249998\, e^{-2\pi m^2/(eE)}+0.111227\, e^{-3\pi m^2/(eE)} \right. \nonumber\\
&\hskip 4cm  \left.  +0.0629846\, e^{-4\pi m^2/(eE)} +\dots \right)
\nonumber
\end{align}
We see that the coefficients of this instanton expansion approach the exact $\frac{1}{k^2}$ factors in (\ref{eq:l1-imag1}).
This demonstrates that our extrapolation from magnetic to electric field is exponentially accurate: it recovers several orders of the exponentially suppressed corrections to the non-perturbative imaginary part of the one-loop effective Lagrangian for an electric background field, using as input 
only 10  terms of the perturbative weak magnetic field expansion. Another way to see this is to plot our improved Pad\'e-Borel transform (\ref{eq:pade-borel1}), shifted slightly from the imaginary axis: Figure \ref{fig:b1-imaginary} shows the resulting poles at integer spacing along the imaginary axis, with residues following the exact $1/k^2$ behavior in (\ref{eq:borel1b}).

\section{The Two-loop Euler-Heisenberg Effective Lagrangian}
\label{sec:twoloop}

\subsection{Exact Results at Two-Loop Order}
\label{sec:2l-exact}

Whereas the one-loop Euler-Heisenberg effective Lagrangian in a constant magnetic field has a simple Borel-Laplace integral representation (\ref{eq:l1}), and 
can be expressed exactly in terms of the  Barnes double gamma function (\ref{eq:barnes}), no such closed-form expressions are known at two-loop for a constant magnetic or electric field background.\footnote{However, exact closed-form expressions and Borel representations are known at two-loop  for a constant {\it self-dual field}, corresponding to the generating function of amplitudes for low-momentum external photons of {\it fixed helicity} \cite{dunne-schubert-sd1,dunne-schubert-sd2,dunne-schubert-sd3}.}
The most explicit representation for a magnetic background field is Ritus's exact double-integral representation \cite{ritus1,ritus2,ritus3,lebedev-ritus}
\begin{equation}
\mathcal{L}^{(2)}\qty(\frac{eB}{m^2})=\frac{B^2}{4}\int_0^\infty \frac{\dd{t}}{t^3}e^{-t\, m^2/(eB)}\qty(J_1+J_2+J_3)
\label{eq:l2a}
\end{equation}
where
\begin{eqnarray}
J_1&=&\frac{2tm^2}{eB}\int_0^1 \frac{\dd{s}}{s(1-s)}\qty[\frac{\cosh(t\, s) \cosh(t(1-s))}{a-b}\ln\frac{a}{b}-t\coth t+\frac{5t^2}{6}s(1-s)]\\
J_2&=&-\int_0^1 \frac{\dd{s}}{s(1-s)}\qty[\frac{c}{(a-b)^2}\ln\frac{a}{b}-\frac{1-b\cosh(t(1-2s))}{b(a-b)}+\frac{b\cosh t+1}{2b^2}-\frac{5t^2}{6}s(1-s)]\\
J_3&=&\qty(1+3\frac{t m^2}{eB}\qty(\ln\left(\frac{t m^2}{eB}\right)+\gamma-\frac{5}{6}))\qty(t\coth t-1-\frac{t^2}{3})
\label{eq:l2b}
\end{eqnarray}
and the functions $a$, $b$ and $c$ are defined as
\begin{eqnarray}
a=\frac{\sinh(t\, s) \sinh(t(1-s))}{t^2\, s(1-s)},\quad b=\frac{\sinh t}{t},\quad c=1-a\cosh(t(1-2s))
\label{eq:l2c}
\end{eqnarray}
Unlike the one-loop case (\ref{eq:l1}), this two-loop expression (\ref{eq:l2a})-(\ref{eq:l2c}) is not directly in Borel form. However, we can expand this as a perturbative weak-field series in $\frac{eB}{m^2}$:
\begin{eqnarray}
\mathcal{L}^{(2)}\qty(\frac{eB}{m^2})\sim
 B^2 \qty(\frac{eB}{m^2})^2\sum_{n=0}^\infty a_n^{(2)}\qty(\frac{eB}{m^2})^{2n}
\quad, \quad e B\ll m^2
\label{eq:l2-weak}
\end{eqnarray}

There is no simple closed-form expression for the coefficients $a_n^{(2)}$. However, the two-loop weak-field coefficients $a_n^{(2)}$ can be generated by a suitable expansion of the integral representation (\ref{eq:l2a}). In \cite{ds-2loop}, 15 terms of such a weak magnetic field expansion were obtained, which was enough perturbative data to argue that the leading large order growth of the two-loop expansion coefficients $a_n^{(2)}$ has exactly the same form as the leading large order growth (\ref{eq:a12}) of the one-loop expansion coefficients $a_n^{(1)}$:
\begin{eqnarray}
a_n^{(2)}&\sim& (-1)^n\frac{\Gamma(2n+2)}{\pi^{2n+2}}+\text{corrections}\quad, \quad n\to\infty
\label{eq:a2-leading}
\end{eqnarray}
In this paper we have pushed the perturbative expansion (\ref{eq:l2-weak}) to much higher order, obtaining 50 terms of the two-loop weak magnetic field expansion. This expansion must be organized appropriately to respect the various subtractions, in order to keep the $s$ integrals finite. Our expansion strategy is described in Appendix \ref{app:ritus}. This new weak-field perturbative data confirms the leading result (\ref{eq:a2-leading}), and furthermore allows analysis of the subleading corrections: see Section \ref{sec:2l-borel} below. The first 25 coefficients $a_n^{(2)}$ are listed in Appendix \ref{app:a2n}, and the first 50 coefficients are listed in an accompanying Supplementary file. This is the perturbative input data on which our subsequent Borel analyses are based.

The leading strong magnetic field behavior at two-loop order is also known \cite{ritus1,ritus2,ritus3} 
\begin{eqnarray}
\mathcal{L}^{(2)} \left(\frac{e B}{m^2}\right) \sim \frac{1}{4}\cdot  \frac{B^2}{2}\qty(\ln\left(\frac{eB}{\pi m^2}\right) -\gamma-\frac{5}{6}+4\zeta(3))
\quad, \quad e B\gg m^2
\label{eq:l2-strong}
\end{eqnarray}
where $\zeta(3)\approx 1.20206$.
As in the one-loop case (\ref{eq:l1-strong}), we identify the two-loop QED beta function coefficient, $\beta_2=1/4$, in the prefactor of the leading logarithmic factor of the strong field limit  (\ref{eq:l2-strong}).

Analytically continuing from a magnetic to an electric background, $B\to i E$, 
the results in (\ref{eq:a2-leading}) and (\ref{eq:l2-strong}) yield the {\it leading} contributions to the non-perturbative imaginary part of the two-loop effective Lagrangian $\mathcal{L}^{(2)}\left(\frac{e E}{m^2}\right)$:
\begin{eqnarray}
{\rm Im}\left[ \mathcal{L}^{(2)}\left(\frac{e E}{m^2}\right) \right]
\sim
\begin{dcases}
 \frac{\pi E^2}{2}\,  e^{-\pi  m^2/(eE)}\quad, \quad e E\ll m^2
\\
~
\\
\beta_2\qty(\frac{\pi}{2})\frac{E^2}{2} \quad,  \quad e E\gg m^2
\end{dcases}
\label{eq:l2-imaginary2}
\end{eqnarray}
These {\it leading} behaviors at two-loop are structurally identical to the leading behaviors at one-loop order (recall (\ref{eq:l1-imag1}) and (\ref{eq:l1-imaginary})). However, we show below that the subleading corrections at two-loop order are very different from the corrections at one-loop order.

\subsection{Borel Analysis at Two-Loop Order}
\label{sec:2l-borel}

Based on the leading factorial growth in (\ref{eq:a2-leading}), and in analogy to the one-loop case analyzed in Section \ref{sec:1l-borel}, we define 
the two-loop (truncated) Borel transform 
\begin{eqnarray}
\mathcal{B}^{(2)}_N(t):=2\sum_{n=0}^{N-1} \frac{a_n^{(2)}}{(2n+1)!}(\pi t)^{2n+1}
\label{eq:borel2}
\end{eqnarray}
\begin{figure}[tb!]
\centering
\includegraphics[scale=.4]{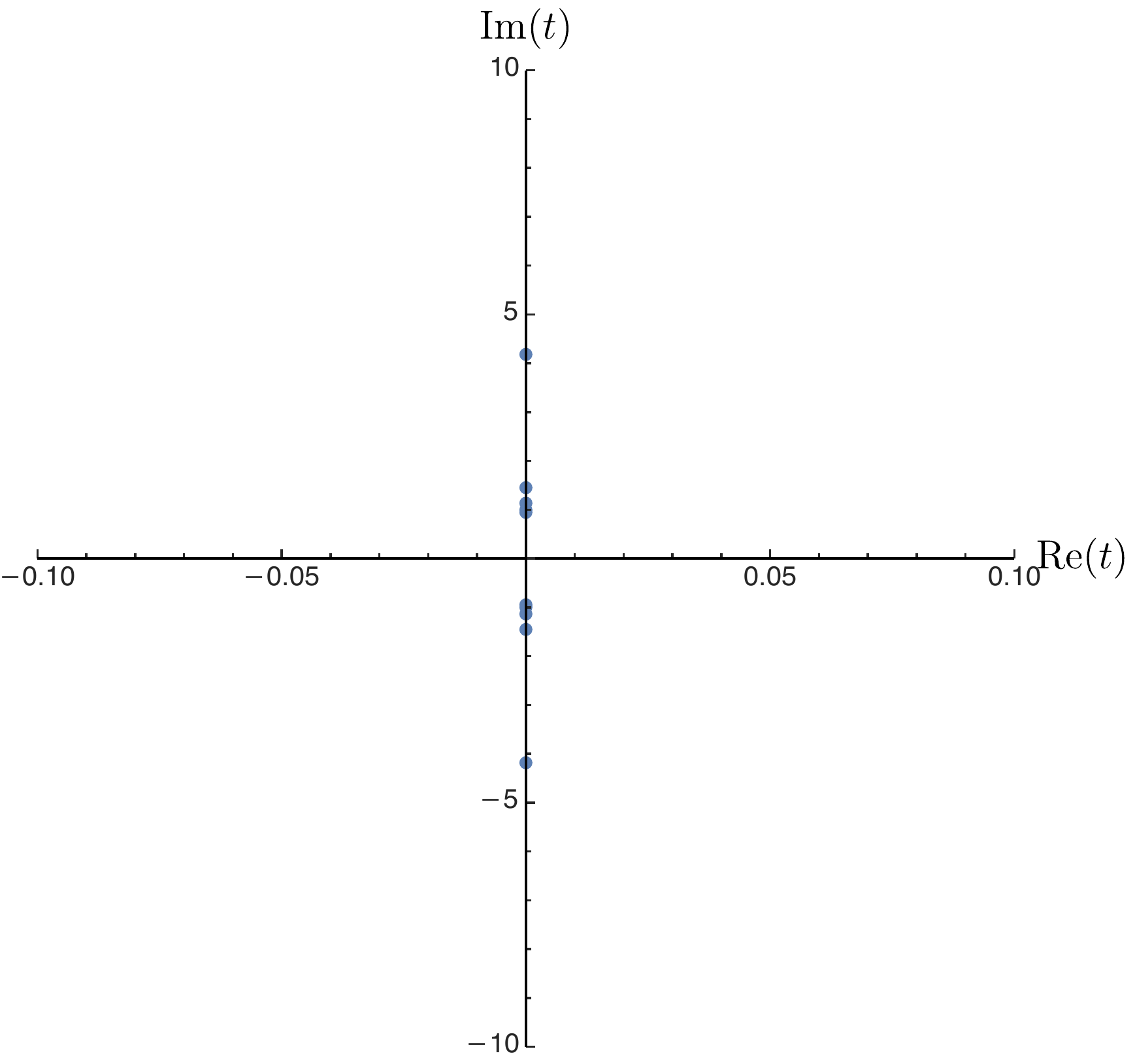}
\includegraphics[scale=.4]{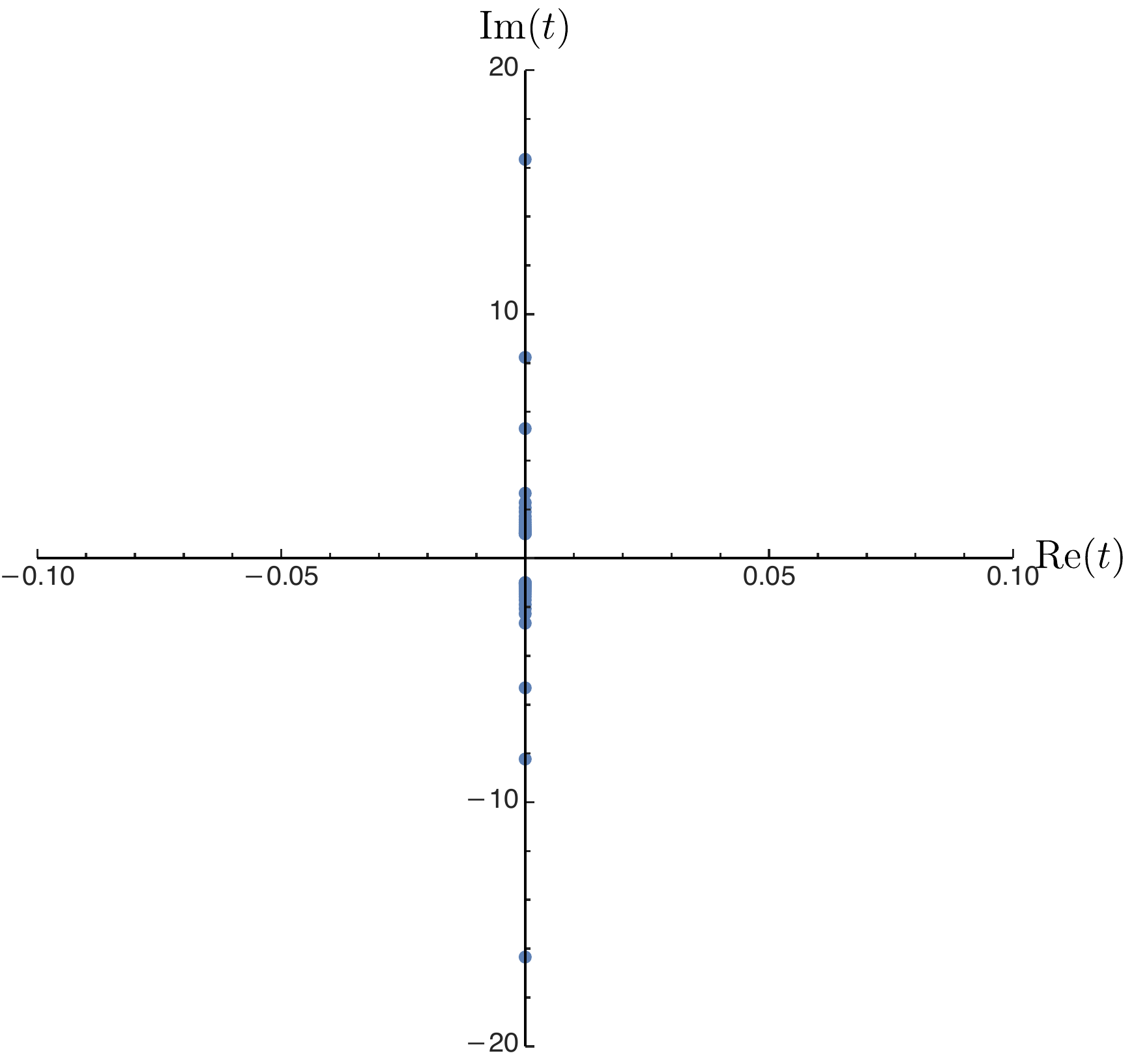}
\caption{Poles of the Pad\'e approximation $\mathcal{PB}^{(2)}_N(t)$ in (\ref{eq:pade-borel2}) for the truncated Borel transform, shown for $N=10$ (left) and $N=50$ (right). Note that the poles appear to be accumulating at $\pm i$. Contrast with one-loop case in Figure \ref{fig:PB1-poles} where the poles are integer-spaced along the imaginary Borel axis.}
\label{fig:PB2-poles}
\end{figure}

Note that we have adopted the same rescaling of $t$ by a factor of $\pi$, as at one loop in (\ref{eq:borel1-n}), because the leading growth of the two-loop coefficients in (\ref{eq:a2-leading}) matches that at one-loop (\ref{eq:a12}). 
Recall also that at one-loop we obtained high-precision analytic continuations by using a modified Pad\'e-Borel transform (\ref{eq:pade-borel1}) that incorporated information about the strong magnetic field behavior of the one-loop effective Lagrangian. Since the strong magnetic field limit in (\ref{eq:l2-strong}) has the same functional form as at one-loop, (\ref{eq:l1-strong}), we adopt the same strategy here. We analytically continue the truncated Borel transform (\ref{eq:borel2}) via a near-diagonal Pad\'e approximant, which encodes the logarithmic strong field behavior (\ref{eq:l2-strong}) at two-loop:
\begin{eqnarray}
\mathcal{PB}^{(2)}_N(t)=\frac{P^{(2)}_N(t)}{Q^{(2)}_{N+1}(t)}
\label{eq:pade-borel2}
\end{eqnarray}
\begin{figure}[tb!]
\centering
\includegraphics[scale=.7]{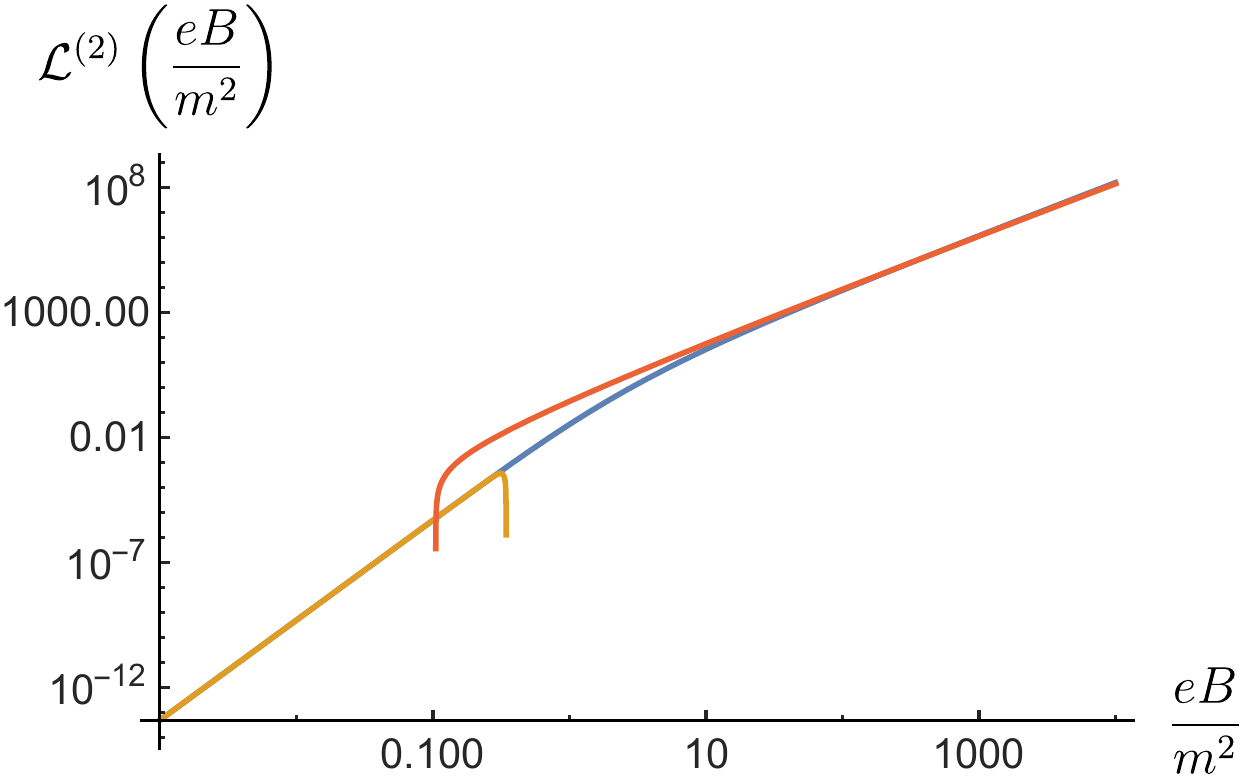}
\caption{The blue curve is a log-log plot of the modified Pad\'e-Borel sum of the truncated weak field expansion in (\ref{eq:pb2}), $\mathcal{L}^{(2)}_N$, plotted here for $N=10$. Compare with the one-loop result plotted in Figure \ref{fig:magnetic1-interpolation}. In contrast to the situation at one loop, there is no exact expression for the Lagrangian at two loop against which to compare the resummation. The gold curve shows the weak field expansion (\ref{eq:l2-weak}), truncated at $N=10$. The red curve shows the leading strong field behavior in (\ref{eq:l2-strong}). Once again, we see that the truncated weak field expansion fails before the Schwinger critical field $eB\approx m^2$, whereas the modified Pad\'e-Borel sum interpolates over many orders of magnitude between the weak-field and strong-field behavior. This plot was made using units in which $e=m^2=1$.}
\label{fig:magnetic2-interpolation}
\end{figure}
\begin{figure}[b!]
\centering
\includegraphics[scale=.7]{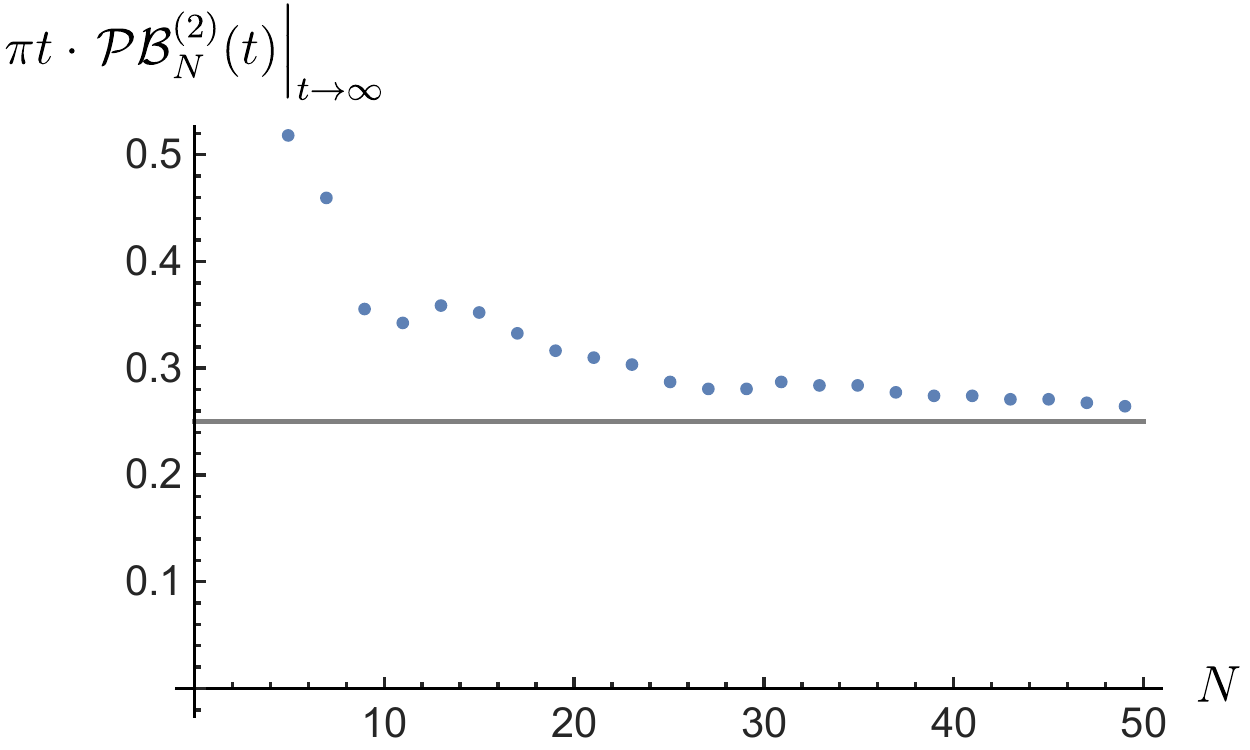}
\caption{
$N$ dependence of the limiting value $\lim\left[\pi t\cdot \mathcal{PB}_N^{(2)}(t)\right]_{t\to\infty}$, for $N$ ranging from 1 to 50, for the near-diagonal Pad\'e approximant of the truncated one-loop Borel transform function in (\ref{eq:pade-borel2}).  The blue dots indicate the values obtained from expanding $\pi t\cdot\mathcal{PB}_N^{(1)}(t)$ about $t\rightarrow\infty$, and which  tend towards the physical value  $\beta_2=1/4$. 
}
\label{fig:beta2}
\end{figure}
Figure \ref{fig:PB2-poles} shows the singularities of this two-loop Pad\'e-Borel transform, $\mathcal{PB}^{(2)}_N(t)$, based on 10 or 50 input coefficients (the first 25 coefficients are listed in Appendix \ref{app:a2n}, and the first 50 coefficients are listed in an accompanying Supplementary file). These plots confirm that the leading singularities are at $t=\pm i$, but they also show that the Borel plane singularity structure is much richer than in the one-loop case, where the singularities are just isolated poles at integer multiples of the leading ones: recall Figure \ref{fig:PB1-poles}. At two-loop, Figure \ref{fig:PB2-poles} suggests that the leading singularities, at $t=\pm i$, appear to be branch points. Recall that a Pad\'e approximant represents branch cuts as lines of poles (interlaced by the Pad\'e zeros) that accumulate to the branch points \cite{Costin:2020hwg,Costin:2020pcj,Stahl}. This novel branch point structure is probed in more detail in Section \ref{sec:power} below.

Given the Pad\'e approximation (\ref{eq:pade-borel2}) to the Borel transform, the approximate two-loop effective Lagrangian for the magnetic background is recovered by the Laplace transform
\begin{eqnarray}
\mathcal{L}_N^{(2)}\! \left(\frac{e B}{m^2}\right)  =\frac{\pi B^2}{2}\int_0^\infty\dd{t}\, e^{-m^2\pi t/(e B)}\,  \mathcal{PB}^{(2)}_N\qty(t)
\label{eq:pb2}
\end{eqnarray}
The quality of this two-loop expression (\ref{eq:pb2}) as an accurate extrapolation of the two-loop effective Lagrangian $\mathcal{L}^{(2)} \!\left(\frac{e B}{m^2}\right)$ from weak magnetic field to strong magnetic field, and from magnetic to electric field, relies on the quality of the analytic continuation of the Borel transform in the Borel plane, here provided by the Pad\'e approximation (\ref{eq:pade-borel2}).

Using only 10 input perturbative coefficients $a_n^{(2)}$, in Figure  \ref{fig:magnetic2-interpolation} we plot the resulting Borel representation $\mathcal{L}_{N}^{(2)}\!\left(\frac{eB}{m^2}\right)$ from (\ref{eq:pb2}). Similar to the one-loop result in Figure \ref{fig:magnetic1-interpolation}, we see that our modified Pad\'e-Borel representation at two-loop also extrapolates accurately over many orders of magnitude from the weak magnetic field to the strong magnetic field regime.
\begin{figure}[tb!]
\centering
\includegraphics[scale=.7]{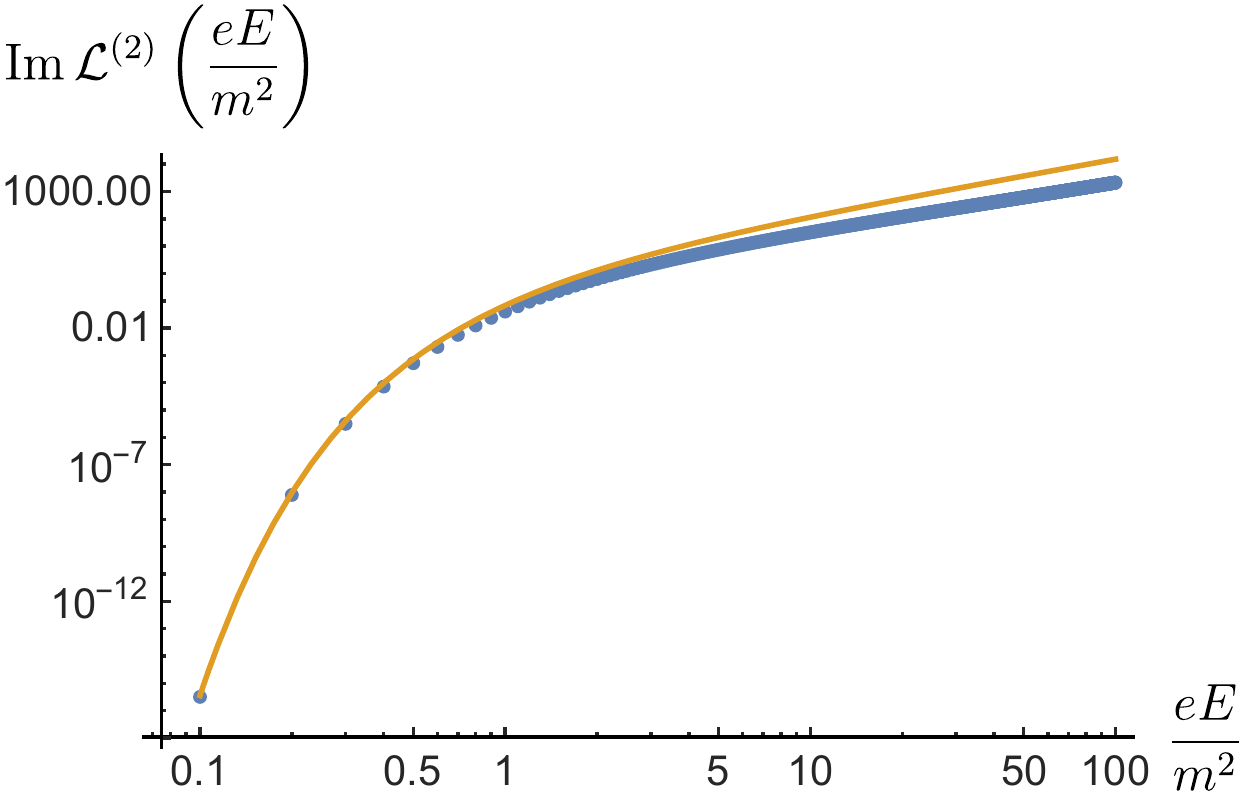}
\caption{A log-log plot of the imaginary part (blue dots) of the two-loop effective Lagrangian in an electric field, calculated using $N=10$ input perturbative terms. The gold curve shows the leading weak-field one-instanton contribution in (\ref{eq:l2-imaginary2}). This plot was made using units in which $e=m^2=1$.}
\label{fig:2l-electric}
\end{figure}
The excellent agreement of this extrapolation to asymptotically large magnetic field can be attributed to the fact that we have constructed our Pad\'e-Borel transform in such a way that it incorporates the form of the known logarithmic behavior (\ref{eq:l2-strong}) of the two-loop effective Lagrangian: the Borel transform, which is generated from an expansion about $t=0$, should be proportional  to $1/(\pi t)$ as $t\to+\infty$. Note that, (as at one-loop) we do not enforce that the coefficient of proportionality be equal to $\beta_2$. Remarkably, once again this fact emerges from our 50 input coefficients, even though these coefficients were generated in the opposite limit near $t=0$.
See Figure \ref{fig:beta2}. 

\begin{figure}[b!]
\centering
\includegraphics[scale=.65]{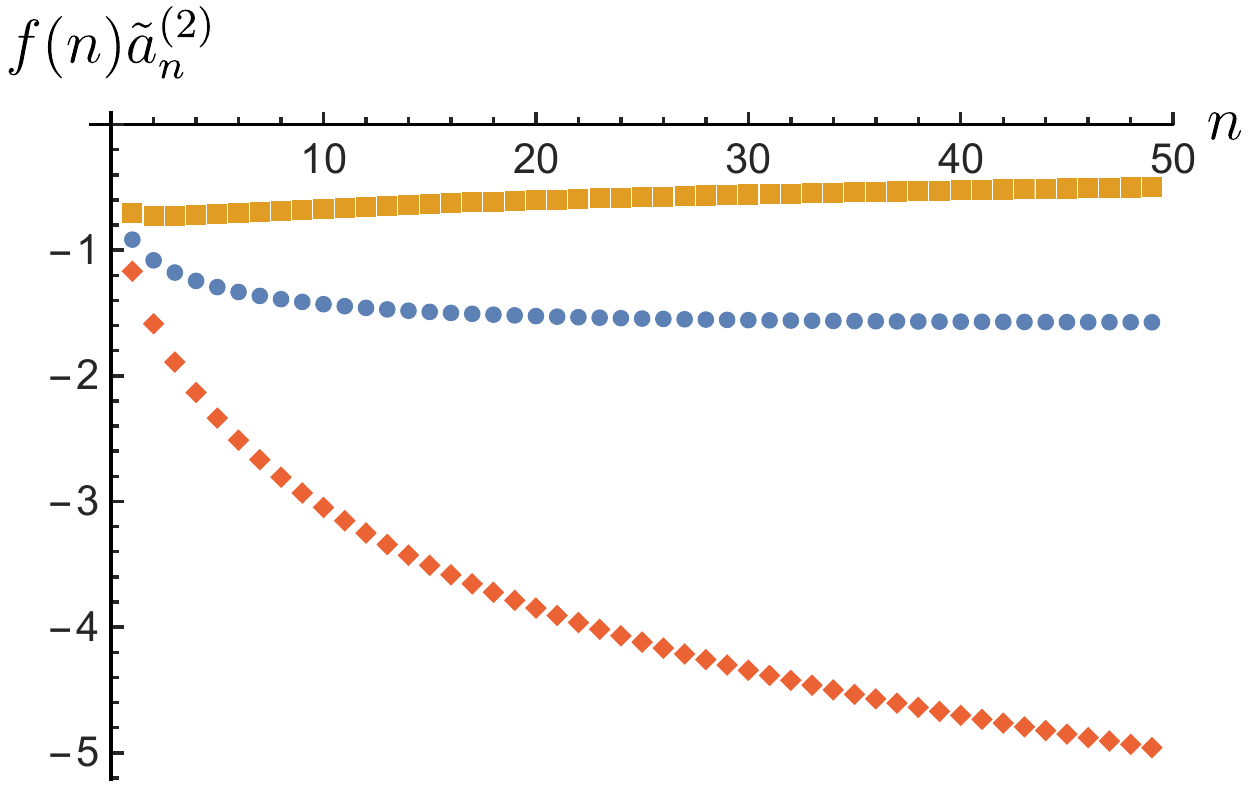}
\caption{The large order behavior of the modified coefficients $\tilde a_n^{(2)}$ defined in (\ref{eq:atilde2}). Different growth rates are shown here: $f(n)=(-1)^n\pi^{2n+2}/\Gamma\qty(2n+\frac{5}{4})$ (blue circles), $f(n)=(-1)^n\pi^{2n+2}/\Gamma\qty(2n+\frac{5}{4}+\frac{1}{4})$ (gold squares), and $f(n)=(-1)^n\pi^{2n+2}/\Gamma\qty(2n+\frac{5}{4}-\frac{1}{4})$ (red diamonds). The form involving $\Gamma(2n+\frac{5}{4})$ is clearly favored.}
\label{fig:a2tilde}
\end{figure}

We can also use the approximate Borel expression (\ref{eq:pb2}) to achieve our second goal: analytic continuation from a magnetic to an electric background, producing an exponentially small non-perturbative imaginary part of $\mathcal{L}_N^{(2)}\! \left(\frac{e E}{m^2}\right)$. The result is shown in Figure \ref{fig:2l-electric}, again showing good agreement over many orders of magnitude of the external field strength. However, there is a clear deviation from this leading weak field contribution even at $eE\approx m^2$, coming from fluctuations about this one instanton term which were not present at one-loop in (\ref{eq:l1-imag1}). This deviation is analyzed in the next Section. See Figure \ref{fig:l2-imag}.

\subsection{Power Law Corrections at Two-Loop Order}
\label{sec:power}

The Pad\'e-Borel pole structure in Figure \ref{fig:PB2-poles} suggests that the leading singularities at two-loop are branch points rather than poles. However, the leading large-order growth in (\ref{eq:a2-leading}) is the same as the one-loop leading large-order growth in (\ref{eq:a1}), and this leading growth is associated with Borel poles. The resolution of this apparent puzzle is that each of the symmetric leading Borel singularities, at $t=\pm i$, is in fact a superposition of a pole and a branch point. This cannot be seen directly from the Pad\'e poles in Figure \ref{fig:PB2-poles}. To resolve this Borel singularity structure, we subtract the exact leading growth behavior and study the remainder, defining modified perturbative weak-field expansion coefficients
\begin{eqnarray}
{\tilde a}^{(2)}_n \equiv a^{(2)}_n -(-1)^n \frac{\Gamma(2n+2)}{\pi^{2n+2}}
\label{eq:atilde2}
\end{eqnarray}
We now analyze the large order behavior of the modified coefficients ${\tilde a}^{(2)}_n$. Ratio tests indicate the following leading 
growth of the ${\tilde a}^{(2)}_n$:
\begin{eqnarray}
{\tilde a}^{(2)}_n \approx \qty(-1.65) \times  (-1)^n  \frac{\Gamma\left(2n+\frac{5}{4}\right)}{\pi^{2n+2}} +\dots 
\label{eq:atilde2-leading}
\end{eqnarray}
See Figure \ref{fig:a2tilde}, where we have adjusted the offset shift, finding the best agreement with the offset $\frac{5}{4}$.

\begin{figure}[h]
\centering
\includegraphics[scale=.7]{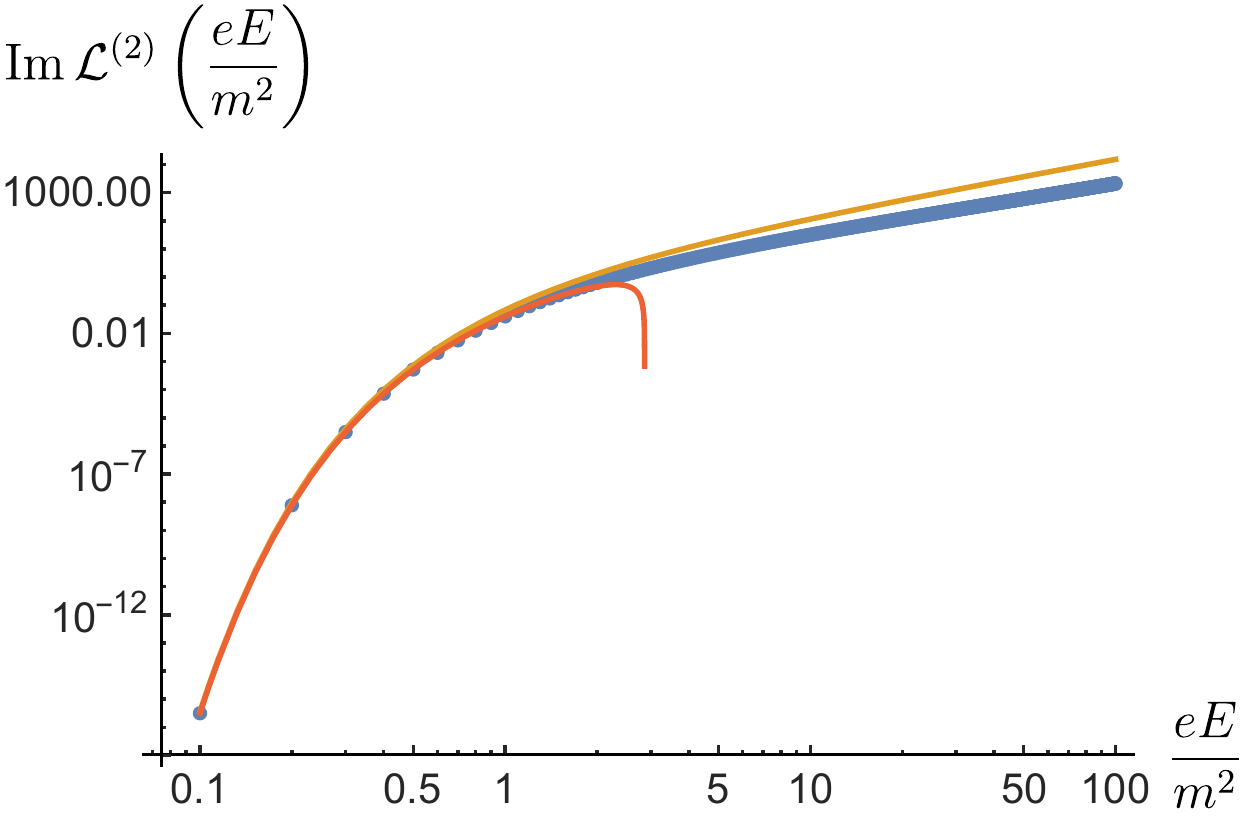}
\caption{The imaginary part of the electric field effective Lagrangian (blue dots), compared with the leading weak-field one-instanton contribution (gold), and the fit including the additional weak-field power-law corrections in (\ref{eq:l2-imaginary2-power-corrections}) (red). This plot was made using units in which $e=m^2=1$.}
\label{fig:l2-imag}
\end{figure}

After analytic continuation to an electric field, this corresponds to the following power-law correction to the imaginary part of the two-loop effective Lagrangian:
\begin{eqnarray}
{\rm Im}\left[{\mathcal L}^{(2)}\left(\frac{e E}{m^2}\right)\right] \sim
\frac{\pi E^2}{2}\, e^{-\pi m^2/(eE)}\bigg[1-1.65\left(\frac{e E}{\pi m^2}\right)^{3/4} 
+ \dots \bigg]
\label{eq:l2-imaginary2-power}
\end{eqnarray}
This form of the leading correction suggests a fluctuation expansion in powers of $\left(\frac{e E}{m^2}\right)^{3/4}$:
\begin{eqnarray}
{\rm Im}\left[{\mathcal L}^{(2)}\left(\frac{e E}{m^2}\right)\right] \sim
\frac{\pi E^2}{2}\, e^{-\pi m^2/(eE)}\bigg[1+d_1 \left(\frac{e E}{\pi m^2}\right)^{3/4} 
 +d_2 \left(\frac{e E}{\pi m^2}\right)^{3/2} +d_3 \left(\frac{e E}{\pi m^2}\right)^{9/4} +\dots \bigg] 
\label{eq:l2-imaginary2-power-corrections}
\end{eqnarray}
In Figure \ref{fig:l2-imag} we plot the result of fitting the imaginary part of the two-loop effective Lagrangian directly from the integral representation. Using the fit interval $\frac{eE}{m^2}\in [10^{-1},1]$ we obtain fit parameters: 
$d_1=-1.65$, $d_2=2.43$, $d_3=-1.94$. 
Figure \ref{fig:l2-imag} illustrates the improved agreement with the successive weak-field corrections. 
This form of the fluctuations about the one-instanton term fills in the first set of missing dots in Ritus's conjectured expression (\ref{eq:l2-imaginary}).

\subsection{Probing the Higher Instanton Terms}
\label{sec:higher}

In fact, the power-law corrections discussed in the previous Section are not the whole story. The weak-field expressions in (\ref{eq:l2-imaginary2}) and (\ref{eq:l2-imaginary2-power-corrections}) only include the effects of the leading Borel singularity at $t=\pm i$: these are  the ``one-instanton'' effects. But we also expect that there should be multi-instanton effects associated with Borel singularities at all integer multiples of the leading ones. These would appear as exponentially small corrections to the large-order growth of the perturbative expansion coefficients $a_n^{(2)}$. 
Therefore, the subleading corrections to the large order growth of the two-loop perturbative expansion coefficients should have the following structural form:
\begin{eqnarray}
a_n^{(2)}&\sim& (-1)^n\frac{\Gamma(2n+2)}{\pi^{2n+2}} \left\{ \big(\text{1+power law corrections}\big) \right. \nonumber\\
&&\left.\qquad\qquad + \big(\text{exponentially small corrections}\big)\times\big(\text{power law corrections}\big)\right\}
\label{eq:a2}
\end{eqnarray}
Even though the {\it leading} large-order growth has the same form as at one-loop, compare (\ref{eq:a12}) and (\ref{eq:a2-leading}), the structure of the corrections is very different: there are power-law corrections followed by much smaller exponentially suppressed corrections, which themselves have power-law corrections. This fact is directly responsible for the novel structure (\ref{eq:l2-imaginary}) of the non-perturbative imaginary part at two-loop order. 
Having studied the structure of the power-law corrections in the previous subsection, we now turn to the exponentially smaller corrections.

\begin{figure}[tb!]
\centering
\includegraphics[scale=.7]{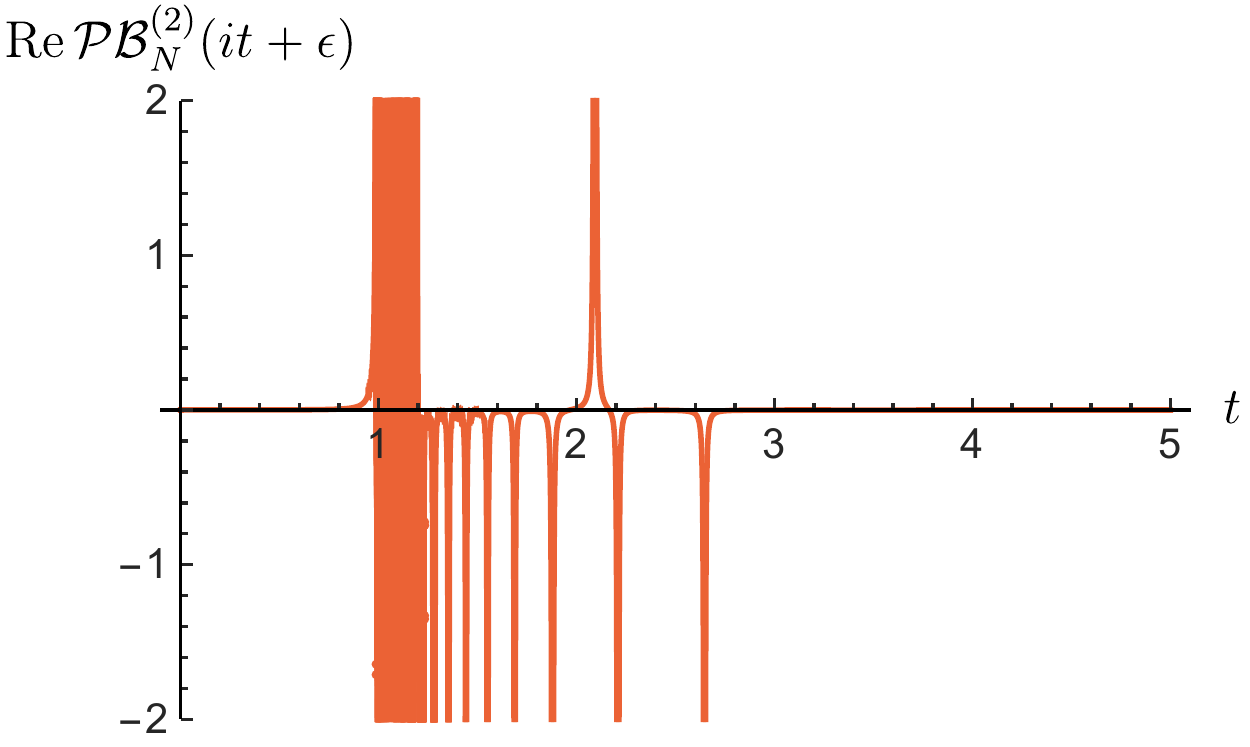}
\caption{Plot of the real part of the Pad\'e-Borel transform, $\Re[\mathcal{PB}_N^{(2)}(it+1/1000)]$ for $N=50$, showing the singularity structure of the Pad\'e-Borel transform along the imaginary axis.  Note that without the conformal map, the accumulation of poles from the Pad\'e-Borel approximation obscures the true singularity structure associated with the physical higher instanton terms. Compare with Figure \ref{fig:pcb2-sub} where the physical multi-instanton Borel singularities at $t=2i$ and $t=3i$ are resolved.}
\label{fig:pb2-sub}
\end{figure}

At one-loop the first few exponentially small corrections can be resolved, see Figure \ref{fig:electric1-subleading}, because there are no power-law corrections (recall (\ref{eq:a12})), but at two-loop it is much more difficult because of the existence of the (much larger) power-law corrections to the leading instanton term. These power-law corrections obscure the exponentially small corrections associated with multi-instantons. This problem can be ameliorated by using more sophisticated Borel techniques, beyond Pad\'e-Borel. Indeed, this problem is directly related to the fact that the Pad\'e-Borel approximation represents the leading branch cut as a line of poles, which therefore obscures the existence of genuine multi-instanton singularities, which also lie on the imaginary axis, and are also expected to be branch points. See Figure \ref{fig:pb2-sub}, which plots the Pad\'e-Borel transform along the imaginary axis: the leading singularity at $t=i$ can be seen, but beyond that one sees coalescing Pad\'e poles that are attempting to represent the branch cut $t\in [i, i\,\infty]$.
This also explains why the multi-instanton Borel singularities are clear at one-loop from the Pad\'e-Borel pole distribution in Figure \ref{fig:PB1-poles} (because they are simple poles), but are not seen directly at two-loop order from the Pad\'e-Borel pole distribution in Figure \ref{fig:PB2-poles} (because they are branch points). Fortunately, there is a simple way to resolve this problem.

\begin{figure}[t]
\centering
\includegraphics[scale=.6]{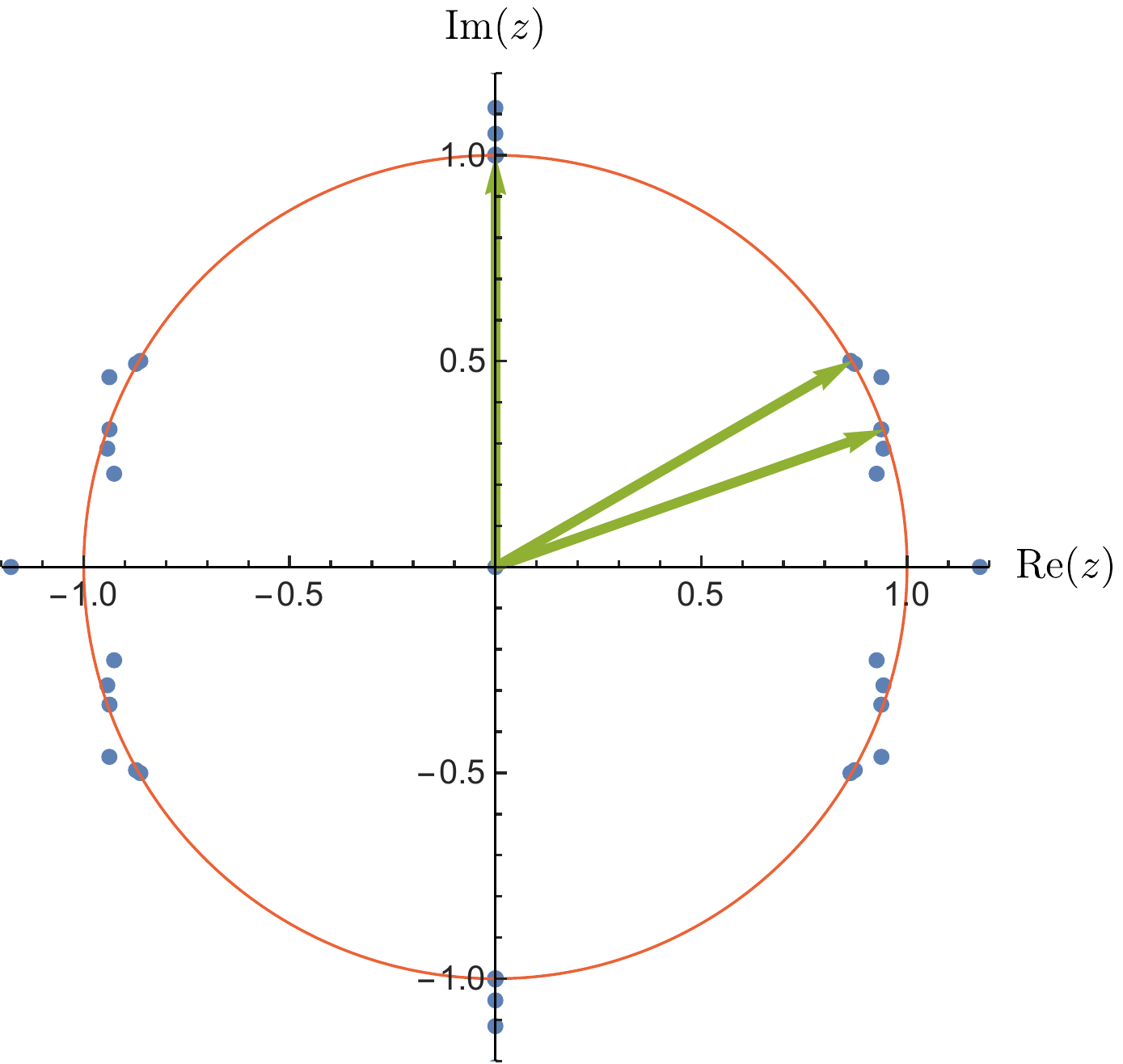}
\caption{Poles of the Pad\'e-Borel approximation in the conformally mapped $z$ plane. With $N=50$ terms, the first three Borel singularities can be resolved as accumulation points of Pad\'e poles located on the unit circle at $\theta=\pm\frac{\pi}{2},\pm\frac{\pi}{6},\pm\arctan\qty(\frac{1}{2\sqrt{2}})$, denoted by the green arrows.}
\label{fig:pcb2-resolved}
\end{figure}

The first step is to confirm that there are indeed integer-repeated Borel singularities, and to determine if they are in fact branch points. This problem can be resolved as follows \cite{Costin:2019xql,Costin:2020hwg,Costin:2020pcj}. We use a conformal map \cite{zinn-book,caliceti,caprini} to map the doubly-cut Borel plane (based on the two symmetric leading branch point singularities at $t=\pm i$) into the unit disk in the conformal $z$ plane.  Specifically, the relevant conformal map for this configuration is: 
\begin{eqnarray}
t= \frac{2z}{1-z^2} \qquad, \qquad z= \frac{t}{1+\sqrt{1+t^2}}
\label{eq:conformal}
\end{eqnarray}
 A re-expansion inside the unit disk to the original order, followed by a Pad\'e approximation within the unit disk, separates subleading branch-points \cite{Costin:2019xql,Costin:2020hwg,Costin:2020pcj}, as shown in Figure \ref{fig:pcb2-resolved}.

The doubly-cut $t$ plane is mapped to the interior of the unit disk, with the edges of the cuts mapped to segments of the boundary, the unit circle. We expand the truncated Borel transform inside the conformal disk, and truncate at the same order as the $t$ expansion:
\begin{eqnarray}
{\mathcal B}_N^{(2)}\left( \frac{2z}{1-z^2}\right)=\sum_{n=0}^{2N-1} b_n^{(2)} \, z^n + \mathcal{O}(z^{2N}) 
\label{eq:cb2}
\end{eqnarray} 
This expansion uniquely defines the coefficients $b_n^{(2)}$. 
By construction, this expansion is convergent within the conformal disk. We then make a near-diagonal Pad\'e approximation, and compute its poles. These singularities are shown in Figure \ref{fig:pcb2-resolved}. 
The poles accumulating to $z=\pm i$ correspond to the leading singularities, since  the conformal map (\ref{eq:conformal}) takes $t=\pm i$ to $z=\pm i$. The next cluster of poles accumulate to $z=\pm e^{\pm i \pi/6}$, which are the conformal map images (on either side of the leading branch cuts) of the two-instanton singularities at $t=\pm 2i$. The third cluster of $z$-plane poles in Figure \ref{fig:pcb2-resolved} accumulate to the images of the three-instanton singularities at $t=\pm 3i$.
Thus the conformal map reveals the existence of integer-repeated higher instanton Borel singularities, and shows that they all have associated branch cuts.

\begin{figure}[h]
\centering
\includegraphics[scale=.65]{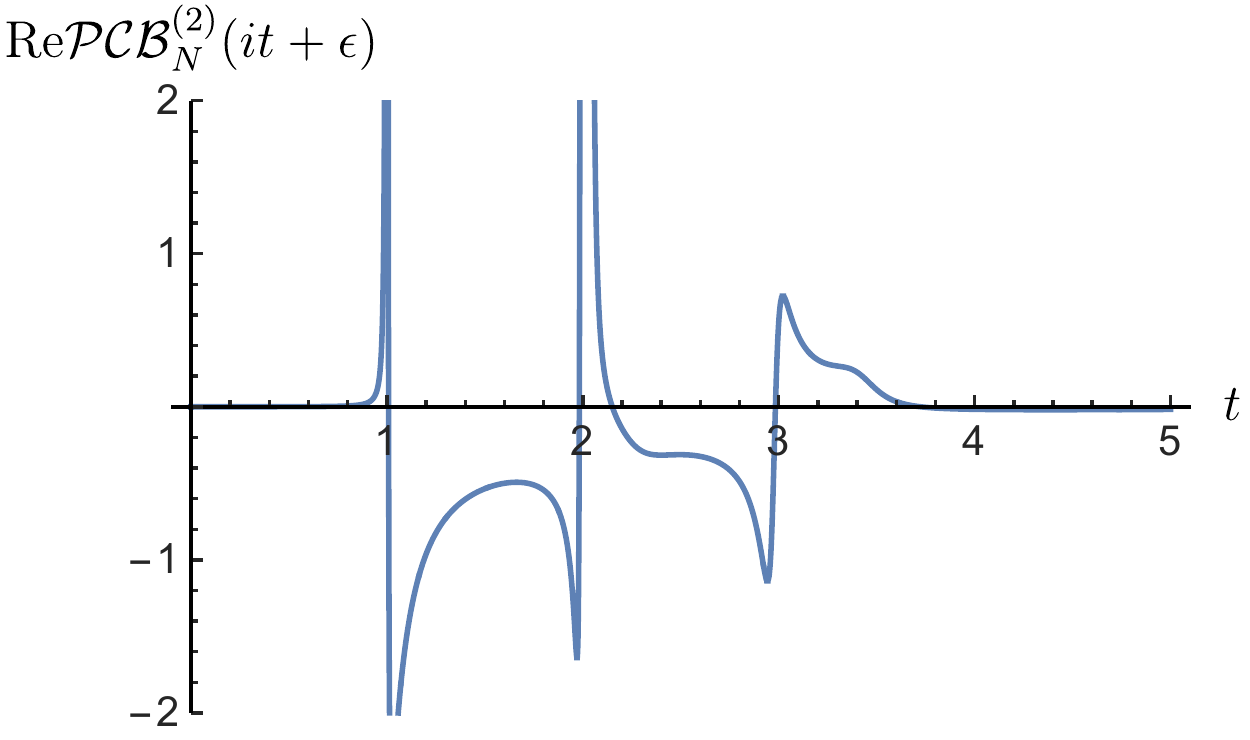}
\caption{Plot of the real part of the Pad\'e-Conformal-Borel transform, $\Re[\mathcal{PCB}_N^{(2)}(it+1/1000)]$ for $N=50$, showing the singularity structure of the conformally mapped Borel transform along the imaginary axis. The plot reveals the existence of higher Borel singularities at multiples of the leading singularity at $t=i$, corresponding to the multi-instanton expansion. 
It also confirms the branch cut nature of the singularity structure at two loop. Compare with Figure \ref{fig:b1-imaginary} at one loop, where the Borel singularities are simple poles, and compare with the analogous plot at two loop, but without the conformal map, in Figure \ref{fig:pb2-sub}.}
\label{fig:pcb2-sub}
\end{figure}

This resolution phenomenon can also be visualized by mapping the Pad\'e approximation within the $z$-plane disk back to the Borel $t$ using the inverse transformation in (\ref{eq:conformal}) \cite{Costin:2019xql,Costin:2020pcj}. This produces the Pad\'e-Conformal-Borel approximation, and this is plotted along the imaginary $t$ axis in Figure \ref{fig:pcb2-sub}. We see that the one-instanton, two-instanton and three-instanton singularities are all resolved.
This should be contrasted with the result of the Pad\'e-Borel approximation, without the conformal map, where nothing beyond the one-instanton singularity can be clearly resolved: recall Figure \ref{fig:pb2-sub}. This failure of the Pad\'e-Borel approximation to resolve higher instanton singularities is a direct consequence of the fact that the Pad\'e approximation represents the leading branch cuts as sequences of poles accumulating to the leading branch points, as seen in Figure \ref{fig:pb2-sub}, and these poles obscure the existence of the genuine physical higher-instanton singularities, which are themselves branch points. The conformal mapping resolves this problem, as can be seen  in Figure \ref{fig:pcb2-sub}.

\begin{figure}[h]
\centering
\includegraphics[scale=.6]{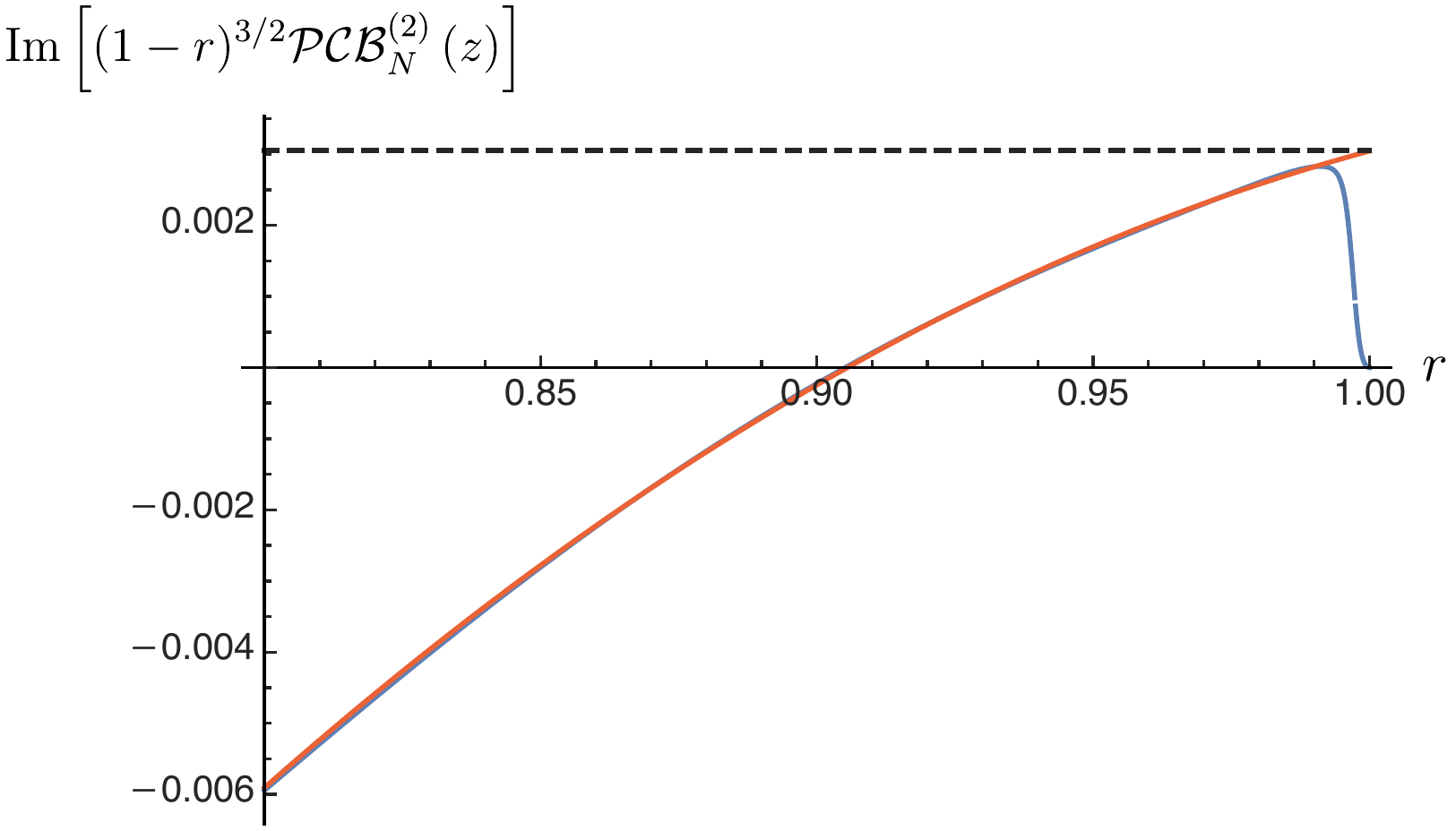}
\caption{Blue: plot of the imaginary part of $(1-r)^{3/2}$ times the Pad\'e-Conformal-Borel transform along the radial line $z=r\, e^{i \pi/6}$ inside the conformal disk, approaching the $z$-plane image of the two-instanton Borel singularity. Red: approximate extrapolation of this plot to the singularity at $z=e^{i \pi/6}$.}
\label{fig:extrap}
\end{figure}

To find the nature of the two-instanton Borel singularity at $t=\pm 2i$, we study the approach to the point $z=e^{i \pi/6}$ in the conformal disk. Writing $z=r\, e^{i \pi/6}$, we assume a power-law behavior $(1-r)^\beta$ for the imaginary part of the Borel transform, and fit a good fit with $\beta=-\frac{3}{2}$. See Figure \ref{fig:extrap}. When mapped back to the Borel $t$ plane, this corresponds to an imaginary part of the two-loop effective Lagrangian, at the two-instanton level, of the form
\begin{eqnarray}
{\rm Im}\left[{\mathcal L}^{(2)}\left(\frac{e E}{m^2}\right)\right]_{ \text{two-instanton}}\sim
 \frac{\pi E^2}{2}\, \sqrt{\frac{m^2}{e E}}\, e^{-2\pi m^2/(eE)}
\label{eq:l2-imaginary22}
\end{eqnarray}
which agrees with the  form of the higher-instanton fluctuation prefactor conjectured by Ritus: recall Eq. (\ref{eq:l2-imaginary}). It is quite remarkable that this doubly-exponentially-suppressed term can be deduced directly from just the first 50 perturbative expansion coefficients of the effective Lagrangian in a magnetic field background.

\section{Conclusions}

We have shown that accurate extrapolations and analytic continuations of the two-loop Euler Heisenberg effective Lagrangian can be recovered from a relatively modest number of terms of the perturbative
weak magnetic field expansion. These perturbative terms are generated from an expansion of Ritus's seminal results for the renormalized two-loop effective Lagrangian in terms of two-parameter integrals \cite{ritus1,ritus2,ritus3,lebedev-ritus}. The new physical effect at two-loop, compared to the well-known one-loop Euler-Heisenberg effective Lagrangian, is that the Borel transform has branch point singularities, rather than just simple poles. These branch points reflect the interactions between virtual particles, and have the effect of producing fluctuation expansions multiplying the terms in the weak-field instanton expansion for the imaginary part of the effective Lagrangian in an electric field background. 
In order to probe these fluctuation corrections we need high precision extrapolations, which we achieve using a combination of  Pad\'e approximations and conformal maps to obtain a sufficiently accurate analytic continuation of the finite
order truncation of the associated Borel transform  \cite{Costin:2020hwg,Costin:2020pcj}. 
We have also incorporated the known physical information about the strong magnetic field limit, which is fixed by the QED beta function, and which determines the functional form of the asymptotic limit of the Borel transform.
In particular, with the input of just 10 terms of the perturbative weak field expansion we find an accurate extrapolation from the weak magnetic field regime to the strong magnetic field regime, over many orders of magnitude. See Figures \ref{fig:magnetic1-interpolation} and \ref{fig:magnetic2-interpolation} for the one-loop and two-loop results, respectively. Using 50 terms of the perturbative weak magnetic field expansion at two-loop order, we analytically continue to an electric field background and obtain new information about the structure of the instanton expansion of the imaginary part of the effective Lagrangian. We  resolve the leading power law correction at the one-instanton level, and also identify the  exponentially further suppressed two-instanton term. 

\begin{figure}[t]
\centering
\includegraphics[scale=0.25]{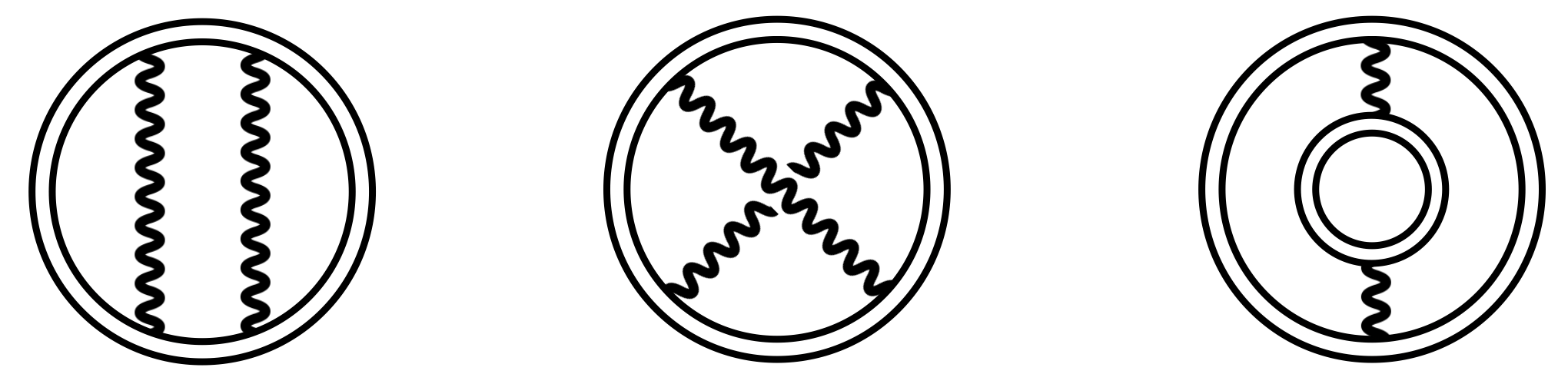}
\caption{The three diagrams which contribute to the Euler-Heisenberg effective Lagrangian at $l=3$ loop order.}
\label{fig:3loop}
\end{figure}

Our analysis was motivated by the question of whether such extrapolations and associated non-perturbative information could be accessible at higher loop order (i.e. higher terms in the expansion (\ref{eq:alpha-exp}) in the fine structure constant), starting not from a closed form multi-parameter integral representation, but from an explicit finite order perturbative expansion. This is because even at three-loop order (see Figure \ref{fig:3loop}) it has so far not been possible to find a parametric integral representation (a $4$-fold parameter integral at three loop) of the Euler Heisenberg effective Lagrangian, even though the exact propagators in a constant background field are known in a relatively simple integral representation form \cite{Huet:2009cy,Huet:2017ydx, Huet:2018ksz}. Our results suggest that an alternative strategy might be more practical: work instead with a perturbative expansion of the propagators, thereby generating a finite-order perturbative expansion of the $l$-loop effective Lagrangian, from which extrapolations to other parametric regimes could be performed. To be practical, such extrapolations must be achievable with a ``reasonable'' amount of perturbative input, and our results suggest that this may indeed be possible. To generate the perturbative expansion at higher loop order, one needs an efficient way to compute the {\it renormalized} effective Lagrangian, for example using the background-field integration-by-parts methods developed in \cite{krasnansky,Dunne:2006sx}.

Certain structural facts are known about the Euler Heisenberg effective Lagrangian at higher loop orders, and these could be used to constrain the higher-loop computations. The exponentiation $e^{\alpha\pi}$ of the leading weak electric corrections to the imaginary part, as in (\ref{all-loop-imag}), leads to a conjecture \cite{Dunne:2004xk} that the leading large order behavior of the perturbative expansion coefficients has the same form for all loop orders $l$:
\begin{eqnarray}
a_n^{(l)}&\sim & (-1)^n \frac{\Gamma(2n+2)}{\pi^{2n+2}}\qquad, \quad n\to\infty\quad, \quad \forall\, l
\label{eq:al}
\end{eqnarray}
Indeed, this conjectured behavior is the reason for the choice of the overall normalization of the perturbative expansion coefficients in (\ref{eq:l-weak}): with this normalization choice we recover the exponential factor $e^{\alpha\pi}$ in (\ref{all-loop-imag}). This conjecture, along with the exponentiation in (\ref{all-loop-imag}), would be interesting to confirm or disprove beyond two-loop order.
Physically, this correspondence is motivated by the interpretation of the mass $m$ appearing in the exponential instanton factor, $e^{-\pi m^2/(e E)}$, as the renormalized physical electron mass \cite{ritus3}. Already at two-loop order, this correspondence between the renormalized mass defined from the real or imaginary part of the effective Lagrangian is sensitive to the finite mass renormalization. The situation at three-loop order is not yet clear \cite{Huet:2009cy,Huet:2017ydx, Huet:2018ksz}, and we hope that the methods described here might provide an alternative approach to shed light on this open question.

\begin{figure}[t]
\centering
\includegraphics[scale=0.275]{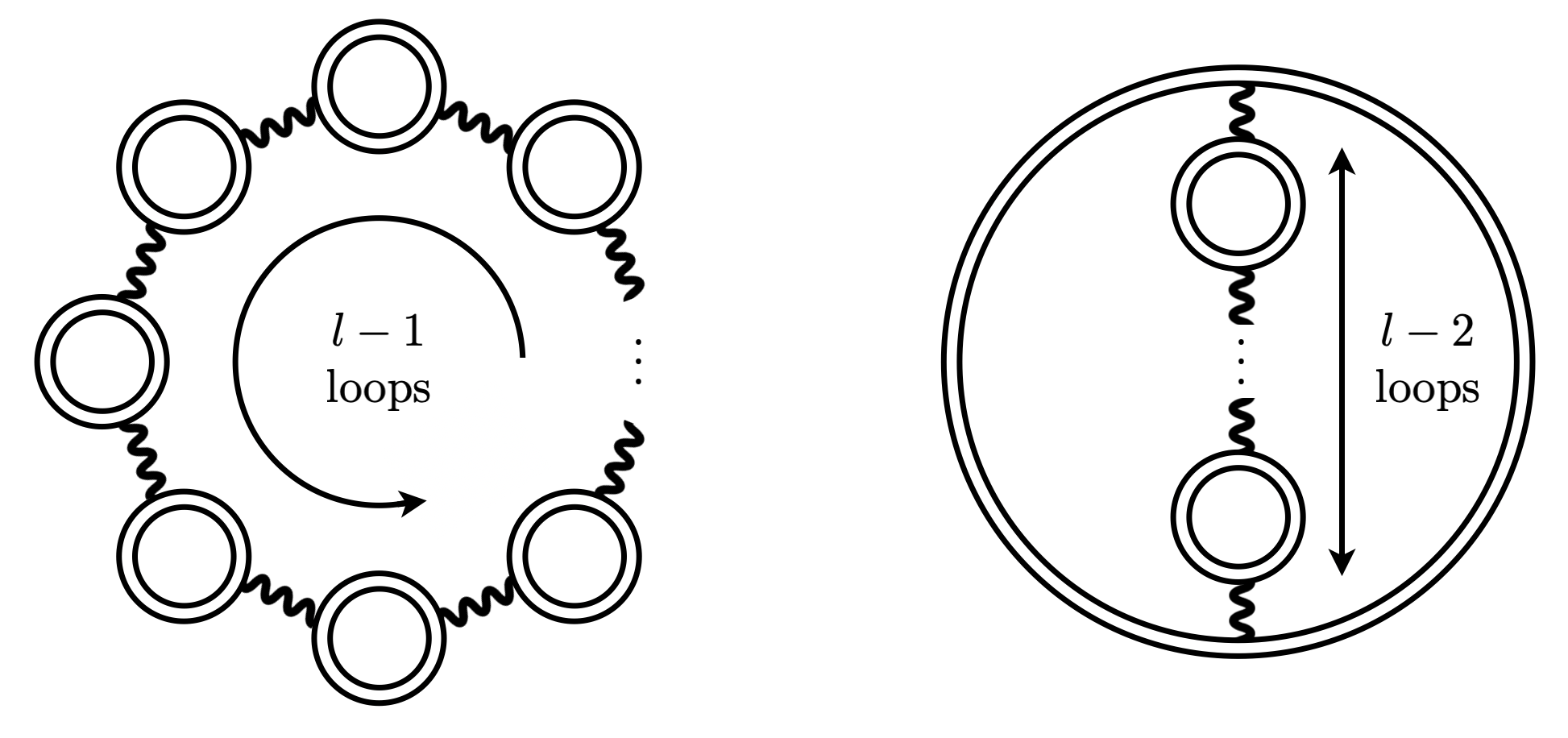}
\caption{Two equivalent views of the irreducible $l$-loop diagram giving the dominant strong-field behavior in (\ref{eq:l-strong}) for the  $l$-loop Euler-Heisenberg effective Lagrangian. There are $(l-1)$ fermion loops, with the double lines denoting fermion propagators in the constant background field, and one overall photon loop.}
\label{fig:renormalon}
\end{figure}

The {\it  leading} strong magnetic field behavior at $l$-loop order (with $l\geq 2$) is also known, arising from the Callan-Symanzik equation in the strong field (or massless) limit \cite{ritus1,ritus2,ritus3}:
\begin{eqnarray}
\mathcal{L}^{(l)}\left(\frac{e B}{m^2}\right) \propto
 \frac{B^2}{2}\left(\ln\left(\frac{eB}{\pi m^2}\right)\right)^{l-1}+\dots 
\quad, \quad e B\gg m^2
\label{eq:l-strong}
\end{eqnarray}
The overall coefficient is expressed in terms of the beta function coefficients up to order $l$. This leading contribution comes from the renormalon-like ``ring diagram'' with $(l-1)$ fermion loops connected in a single ring by $(l-1)$ photon propagators. See Figure \ref{fig:renormalon}. This general fact could be used to constrain the asymptotic behavior of the Borel transform at $l$-loop order.\footnote{It would be interesting to apply these perturbative Borel methods also to the reducible diagrams studied in \cite{Gies:2016yaa,Karbstein:2019wmj}.} Deeper understanding of the strong field limit at higher order in the fine structure constant $\alpha$ may also shed light on the computation of scattering amplitudes in strong background fields, in particular those associated with ultra-intense lasers \cite{DiPiazza:2011tq,Meuren:2020nbw}.
For example, seminal work by Ritus and Narozhnyi has made predictions for the resulting structure at higher loop order for the special case where the background laser field is represented as a constant crossed field \cite{Narozhnyi:1979at,Narozhnyi:1980dc,Morozov:1981pw,Fedotov:2016afw,Podszus:2018hnz,Ilderton:2019kqp,Mironov:2020gbi}.

\section{Appendix: Ritus's Exact Double-Integral Representation}
\label{app:ritus}

The two-loop Euler-Heisenberg effective Lagrangian can be written as the double integral 
\begin{align}
\mathcal{L}^{(2)}\qty(\frac{eB}{m^2})&=\frac{B^2}{4}\int_0^\infty \frac{\dd{t}}{t^3}e^{-tm^2/(eB)}\qty(J_1+J_2+J_3)
\end{align}

with
\begin{eqnarray}
J_1&=&\frac{2tm^2}{eB}\int_0^1 \frac{\dd{s}}{s(1-s)}\qty[\frac{\cosh ts \cosh t(1-s)}{a-b}\ln\frac{a}{b}-t\coth t+\frac{5t^2}{6}s(1-s)]\\
J_2&=&-\int_0^1 \frac{\dd{s}}{s(1-s)}\qty[\frac{c}{(a-b)^2}\ln\frac{a}{b}-\frac{1-b\cosh t(1-2s)}{b(a-b)}+\frac{b\cosh t+1}{2b^2}-\frac{5t^2}{6}s(1-s)]\\
J_3&=&\qty(1+3\frac{tm^2}{eB}\qty(\ln\left(\frac{tm^2}{eB}\right)+\gamma-\frac{5}{6}))\qty(t\coth t-1-\frac{t^2}{3})
\end{eqnarray}

and
\begin{align}
a=\frac{\sinh ts \sinh t(1-s)}{t^2s(1-s)},\quad b=\frac{\sinh t}{t},\quad c=1-a\cosh t(1-2s)
\end{align}

To generate a weak magnetic field expansion, for $J_1$ and $J_2$ we expand each term in the above expressions at small $t$ to order $\mathcal{O}(t^{2n})$. This is straightforward for most terms, with
\begin{align}
\ln\frac{a}{b}=\sum_{k=0}^n \frac{2^{2k}\qty(s^{2k}+(1-s)^{2k}-1)}{2k(2k)!}t^{2k}
\end{align}

For the factors $\qty(a-b)^{-p}$, we first expand
\begin{align}
a-b=t^2\sum_{k=0}^{n-1}\qty[\frac{1}{(2k+4)!}\frac{1-(1-2s)^{2k+4}}{2s(1-s)}-\frac{1}{(2k+3)!}]t^{2k}=t^2\sum_{k=0}^{n-1}A_k t^{2k}
\end{align}

Then, the coefficients of this Taylor series raised to an arbitrary negative power can be generated recursively
\begin{align}
(a-b)^{-p}=\frac{1}{t^{2p}}\sum_{k=0}^{n+p}A^{(-p)}_k t^{2k}
\end{align}

where (for $A_0\neq 0$)
\begin{align}
A_0^{(-p)}&=\frac{1}{A_0^p}\\
A_{k}^{(-p)}&=\frac{1}{kA_0}\sum_{\ell=1}^k\qty[\ell(1-p)-k]A_\ell A^{(-p)}_{k-\ell},\quad k\geq 1
\end{align}

We then combine these expansions using a discrete convolution
\begin{gather}
\qty(\sum_{k=0}^n a_k t^{2k})\qty(\sum_{k=0}^n b_k t^{2k})=\sum_{k=0}^n c_k t^{2k}\\
c_k=\sum_{\ell=0}^k a_\ell b_{k-\ell}
\end{gather}

Although certain terms in $J_1$ and $J_2$ 
look divergent with respect to the $s$ integral, they are exactly canceled by the expansion of  other terms in the integrand. In addition, each of the remaining terms contains a factor of $s(1-s)$, leaving well-defined integrals which result in polynomials in $t$. For $J_3$, there is no $s$ integral to perform, and so we can just expand the entire expression at small $t$ to $\mathcal{O}(t^{2n})$, yielding polynomials in $t$ and polynomials multiplied by $\ln \frac{tm^2}{eB}$. Now, all the $t$ integrals can be performed, where we can make use of the result
\begin{equation}
\int_0^\infty \dd{t} e^{-m^2t/(eB)} t^{2n} \ln\left(\frac{tm^2}{eB}\right) =\qty(\frac{eB}{m^2})^{2n+1}\Gamma(2n+1)\psi^{(0)}(2n+1),\quad n>-\frac{1}{2}
\end{equation}
with $\psi^{(0)}(x)$ the digamma function. With this algorithm, we were able to generate fifty coefficients in the weak magnetic field expansion of $\mathcal{L}^{(2)}\qty(\frac{eB}{m^2})$, whereas previous analysis only obtained fifteen coefficients \cite{ds-2loop}. The first 25 weak magnetic field expansion coefficients of the two loop Euler-Heisenberg Lagrangian can be found in Appendix \ref{app:a2n}, and the first 50 coefficients are contained in an accompanying Supplementary file.

\vfill

\section{Appendix: Coefficients of the Two-Loop Weak Magnetic Field Expansion}
\label{app:a2n}

\begin{table}[H]
\centering
\renewcommand{\arraystretch}{2.5}
\begin{tabular}{l|ll}
$n$ & ~ & $ a_n^{(2)}$\\
\hline $0$&~&$\displaystyle\frac{4}{81}$\\

$1$&~&$\displaystyle-\frac{1219}{32400}$\\

$2$&~&$\displaystyle\frac{33827}{396900}$\\

$3$&~&$\displaystyle-\frac{98923}{255150}$\\

$4$&~&$\displaystyle\frac{532455472}{180093375}$\\

$5$&~&$\displaystyle-\frac{10457103346}{307432125}$\\

$6$&~&$\displaystyle\frac{8280132424}{15035625}$\\

$7$&~&$\displaystyle-\frac{1322128891875104}{110718132525}$\\

$8$&~&$\displaystyle\frac{216079874926085406464}{646561329789375}$\\

$9$&~&$\displaystyle-\frac{399159024629278987264}{34029543673125}$\\

$10$&~&$\displaystyle\frac{118088344460059990083104768}{234021171840080625}$\\

$11$&~&$\displaystyle-\frac{10979187935861311150383104}{420447667696875}$\\

$12$&~&$\displaystyle\frac{18653011055800029685394341888}{11654809348557375}$\\

$13$&~&$\displaystyle-\frac{1635567621127339081554728370176}{14265121494999375}$\\

$14$&~&$\displaystyle\frac{214273566180161138094998308030185472}{22571190592506430875}$\\

$15$&~&$\displaystyle-\frac{25109740072479538970451463102102765568}{27910612022991823125}$\\

$16$&~&$\displaystyle\frac{764034086290611781552561173252582378831872}{7896673339814104720875}$\\

$17$&~&$\displaystyle-\frac{5323656802968440248000341338657607172260626432}{454224613538046611885625}$\\

$18$&~&$\displaystyle\frac{214511358602073886111889879319593742860030050304}{135039749970770614344375}$\\

$20$&~&$\displaystyle\frac{266462893626229350602869353007174008140288948032879198208}{6672849284578939554478010625}$\\

$21$&~&$\displaystyle-\frac{2437894565001775347497238208128324910365652981172597161984}{332534017503933864841096875}$\\

$22$&~&$\displaystyle\frac{102630737774414418996011603373560456263255744267423025320689664}{69566951104903323342538824375}$\\

$23$&~&$\displaystyle-\frac{264983617781753579377541150234960277977765158821145927518255054848}{817620826815723148073039559375}$\\

$24$&~&$\displaystyle\frac{34888103723767870212111639769661161527345864219016664926050448310272}{450525761714786224448409553125}$\\

\end{tabular}
\end{table}

\acknowledgements

We thank Ovidiu Costin for helpful discussions. This material is based upon work supported by the U.S. Department of Energy, Office of Science, Office of High Energy Physics under Award Number DE-SC0010339, and by the National Science Foundation under Grant DHE-1747453.

\end{document}